\documentclass[iop,apj,numberedappendix]{emulateapj}

\usepackage{acronym}
\usepackage{booktabs}
\usepackage{dcolumn}
\usepackage{graphicx}
\usepackage{amsmath,amsfonts,amssymb}
\usepackage{mathrsfs}
\usepackage[normalem]{ulem}
\usepackage[usenames,dvipsnames,svgnames,table]{xcolor}
\usepackage[caption=false]{subfig}

\usepackage[breaklinks,colorlinks,citecolor=blue]{hyperref}

\newcolumntype{d}[1]{D{.}{.}{#1}}

\usepackage{natbib}
\newcommand{\refeq}[1]{Eq.~(\ref{eq:#1})}          
\newcommand{\refeqs}[2]{Eqs.~(\ref{eq:#1})--(\ref{eq:#2})}          
\newcommand{\reffig}[1]{Fig.~\ref{fig:#1}}          
\newcommand{\refsec}[1]{Sec.~\ref{sec:#1}}
\newcommand{\refapp}[1]{App.~\ref{app:#1}}

\newcommand{\msun}{\, {\rm M}_\odot }

\def\bfd{\boldsymbol{d}}

\def\bfalp{\boldsymbol{\alpha}}

\def\bfx{\mathbf{x}}
\def\bfy{\mathbf{y}}
\def\bfA{\mathbf{A}}
\def\bfd{\mathbf{d}}

\usepackage{color}

\newcommand{\sasha}[1]{}

\newcommand{\sk}[1]{}

\newcommand{\be}{\begin{equation}}
\newcommand{\ee}{\end{equation}}
\newcommand{\ba}{\begin{eqnarray}}
\newcommand{\ea}{\end{eqnarray}}
\newcommand{\en}{\nonumber\\}

\begin{document}

\title{Probing dark matter subhalos in galaxy clusters using highly magnified stars}

\author{Liang Dai\altaffilmark{1,4,*}, Tejaswi Venumadhav\altaffilmark{1}, Alexander A. Kaurov\altaffilmark{1}, and Jordi Miralda-Escud\'{e}\altaffilmark{2,3}}

\altaffiltext{1}{Institute for Advanced Study, 1 Einstein Drive,
Princeton, NJ 08540, USA}
\altaffiltext{2}{Instituci\'o Catalana de Recerca i Estudis Avan\c cats, Barcelona, Catalonia}
\altaffiltext{3}{Institut de Ci\`encies del Cosmos, Universitat de Barcelona (IEEC-UB), Barcelona, Catalonia}
\altaffiltext{4}{NASA Einstein Fellow.}
\altaffiltext{*}{Electronic address: \href{mailto:ldai@ias.edu}{ldai@ias.edu}}

\begin{abstract}

Luminous stars in background galaxies straddling the lensing caustic of a foreground galaxy cluster can be individually detected due to extreme magnification factors of $\sim 10^2$--$10^3$, as recently observed in deep {\it HST} images. We propose a direct method to probe the presence of dark matter subhalos in galaxy clusters by measuring the astrometric perturbation they induce on the image positions of magnified stars or bright clumps: lensing by subhalos breaks the symmetry of a smooth critical curve, traced by the midpoints of close image pairs. For the giant arc at $z = 0.725$ behind the lensing cluster Abell 370 at $z = 0.375$, a promising target for detecting image pairs of stars, we find that subhalos of masses in the range $10^6$--$10^8\msun$ with the abundance predicted in the cold dark matter theory should typically imprint astrometric distortions at the level of $20$--$80\,{\rm mas}$. We estimate that $\sim 10\,$hr integrations with {\it JWST} at $\sim 1$--$3\,\mu{\rm m}$ may uncover several magnified stars whose image doublets will reveal the subhalo-induced structures of the critical curve. This method can probe a dynamic range in the subhalo to cluster halo mass ratio $m/M \sim 10^{-7}$--$10^{-9}$, thereby placing new constraints on the nature of dark matter.

\end{abstract}


\section{introduction}
\label{sec:intro}

The Cold Dark Matter (CDM) paradigm predicts that gravitational collapse of primordial density fluctuations produces bound dark matter halos with a wide range of masses~\citep{1981ApJ...250..423D, blumenthal1982galaxy, blumenthal1984formation, 1985ApJ...292..371D}, the smallest of which may be as tiny as $10^{-6} M_\odot$ \citep{PhysRevLett.97.031301,diemand2005earth}. Small halos merge to assemble massive halos and create a hierarchical structure all the way up to the scale of galaxy clusters, with masses $\sim 10^{15} M_\odot$~\citep{press1974formation, white1978core}.

Observational probes of dark matter halos over a broad mass spectrum are crucial for testing the CDM paradigm. Several hypotheses for the nature of the dark matter predict a deficit of halos below characteristic mass scales: these include warm dark matter~\citep{PhysRevD.71.063534, colin2000substructure, bode2001halo}, dark matter with a macroscopic de Broglie wavelength~\citep{ PhysRevD.28.1243, PhysRevD.50.3650, PhysRevLett.85.1158, goodman2000repulsive, peebles2000fluid, PhysRevD.95.043541}, and dark matter as compact objects~\citep{carr1974black, meszaros1974behaviour, Carr:1975qj}.

When available, gravitational lensing is the most direct probe of dark matter halos in the absence of stars. Both strong and weak lensing have been used to measure the masses of halos hosting galaxy clusters~\citep{2013SSRv..177...75H} and galaxies~\citep{2010ARA&A..48...87T, 2015IAUS..311...86M}. Measurements of the orbital motions of stars and gas have also enabled us to reconstruct the dark matter distribution in the inner parts of galaxy halos \citep{2001ARA&A..39..137S, 2015NatPh..11..245I}. Detecting halos on sub-galactic scales ($\lesssim 10^{10}\,M_\odot$) has proven challenging because these typically contain little luminous matter~\citep{2008MNRAS.390..920O}. Stellar dynamics measurements suggest that the ultra-faint satellites of the Milky Way could be held together by dark matter halos of sub-galactic masses $10^6$-$10^8\,M_\odot$~\citep{2010arXiv1009.4505B}, although the initial extent and total mass of these halos is unclear. Subhalos are also expected to leave characteristic imprints in stellar streams~\citep{johnston2002lumpy, ibata2002uncovering, carlberg2009star}.

Strong gravitational lensing offers powerful tools to probe the distribution of dark matter within lensing halos. A classic application is to use magnification ratios of multiple images of background quasars or galaxies to detect subhalos in galaxy halos~\citep{Mao:1997ek, metcalf2001compound, dalal2002direct, 2017MNRAS.471.2224N, vegetti2010detection, vegetti2012gravitational, Cyr-Racine:2015jwa}. Dark matter substructure can also be detectable through distortions in the surface brightness patterns of lensed giant arcs seen at submillimeter~\citep{hezaveh2013dark, hezaveh2016detection, asadi2017probing} or optical/infrared wavelengths~\citep{2017JCAP...05..037B}. Together, these methods have yielded many recent detections of subhalos with masses of $10^8$-$10^9\,M_\odot$. Lowering the mass detection limit would allow comparisons with other probes of sub-galactic scale halos, which suffer from very different systematics. 

This paper proposes a novel probe of subhalos in lensing galaxy clusters. Our method requires two ingredients: a background galaxy that straddles the lensing caustic of a cluster, and a number of luminous stars (or compact light clumps) detected close to the critical curve. The image pairs of these stars (or light clumps) trace the critical curve and measure its degree of irregularity as affected by the subhalos.
 
Highly magnified images naturally occur in pairs straddling the critical curve of the cluster. For a smooth lens mass distribution, the lensing geometry takes the form of a fold catastrophe~\citep{schneider1999gravitational}, with images lying on symmetric positions, and the radius of curvature of the critical curve is comparable to the cluster's Einstein radius, $\theta_{\rm E} \sim 20''$. When subhalos are introduced, the critical curve develops small-scale distortions, with perturbations that are dramatically enhanced in the vicinity of the critical curve even for a small subhalo abundance, analogously to the microlensing effects of intracluster stars~\citep{Venumadhav:2017ttg}.

For typical parameters of cluster lenses, stars located $\sim 10 \, {\rm pc}$ from the caustic are magnified by factors of $\sim 10^2$--$10^3$. The most luminous stars ($L \gtrsim 10^5\,L_\odot$) can be individually detectable at optical/IR wavelengths from cosmological redshifts $z_s \sim 1$~\citep{miralda1991magnification}. Recently, a first detection was reported in Hubble Space Telescope ({\it HST}) images of the strong lensing cluster MACSJ1149~\citep{2017arXiv170610279K}. We show that the giant arc in Abell 370~\citep[$z_s = 0.725$; see][]{2010MNRAS.402L..44R} is a promising target for detecting the images of several stars with the upcoming James Webb Space Telescope ({\it JWST}). If the separation between image pairs can be resolved, they can be robustly identified and used as tracers of the critical curve. For the test case in Abell 370, we show that at the sensitivity of {\em JWST}, the expected angular separation between detectable image-pairs of magnified stars is $\sim 0.1''$, which is above the astrometric resolution. We propose to search for this distortion by performing astrometry on image pairs. In our test case, the typical size of the astrometric distortion is $\sim 20-100\,{\rm mas}$ due to subhalos with masses of $\sim 10^6$--$10^8\msun$.

Microlensing by intracluster stars causes short-time variability in the magnified images of stars that allows for their identification~\citep{Diego:2017drh, Venumadhav:2017ttg, Oguri:2017ock, Windhorst:2018wft}. However, in this paper we are interested in using images of stars or of other bright clumps in the lensed galaxies as a probe of the shape of the critical curve, for which microlensing variability is not crucial. Accurate astrometry can be done with images of any source (stellar clusters, HII regions or irregularities in the lensed galaxy surface brightness) that is compact enough and close to the caustic.

The paper is organized as follows. In \refsec{lensing}, we first review the caustic crossing phenomenon in a smooth fold model without substructure, and then discuss lensing perturbations from subhalos and the impact of microlensing by intracluster stars. In \refsec{sim}, we show that the giant arc in the strong lensing cluster Abell 370 is particularly promising for detecting several caustic crossing luminous stars, and present simulated lensing signatures of subhalos imprinted in the image positions within the arc. Finally, we summarize the results and discuss observational prospects for our method with optical/infrared telescopes in \refsec{concl}, highlighting the scientific potential of the forthcoming {\it JWST}. For interested readers, we collect some of the technical information in the Appendices. Appendix \ref{app:shabund} details our modeling of the subhalo abundance based on existing N-body simulations. Appendix \ref{app:shlen} presents the analytical model used for the lensing effect of individual subhalos. Appendix \ref{app:raytrace} briefly explains the ray-shooting code used for the lensing simulations. 

Throughout this paper, we adopt the {\it Planck} best-fit cosmological parameters of \cite{ade2016planck}. We work with geometrical optics, which is an excellent approximation for lensing at UV and infrared wavelengths.

\section{Lensing of caustic crossing stars}
\label{sec:lensing}

The method laid out in this paper relies on the behavior of lens models near singularities, the most common of which is the fold. After briefly reviewing the fold, we proceed to discuss the perturbing effect of substructure. 

\subsection{Lensing without subhalos: fold model}
\label{sec:fold}

We follow the notation of \cite{Venumadhav:2017ttg} for the lens mapping $\bfy = \bfy(\bfx)$ from the image position $\bfx$ to the source position $\bfy$, in angular units. The Jacobian matrix of the map is $\bfA(\bfx) = \partial \bfy/\partial \bfx$. In terms of this Jacobian matrix, the (signed) magnification factor $\mu(\bfx)$ is the inverse of the determinant, i.e., $\mu(\bfx) = 1/\det[\bfA(\bfx)]$, the convergence $\kappa(\bfx)$ is related to the trace by $\kappa(\bfx) = 1 - ({\rm tr}[\bfA(\bfx)]/2)$, and the lensing shear is the trace-free and symmetric part. 

\begin{figure}[t]
  \begin{center}
    \includegraphics[scale=0.75]{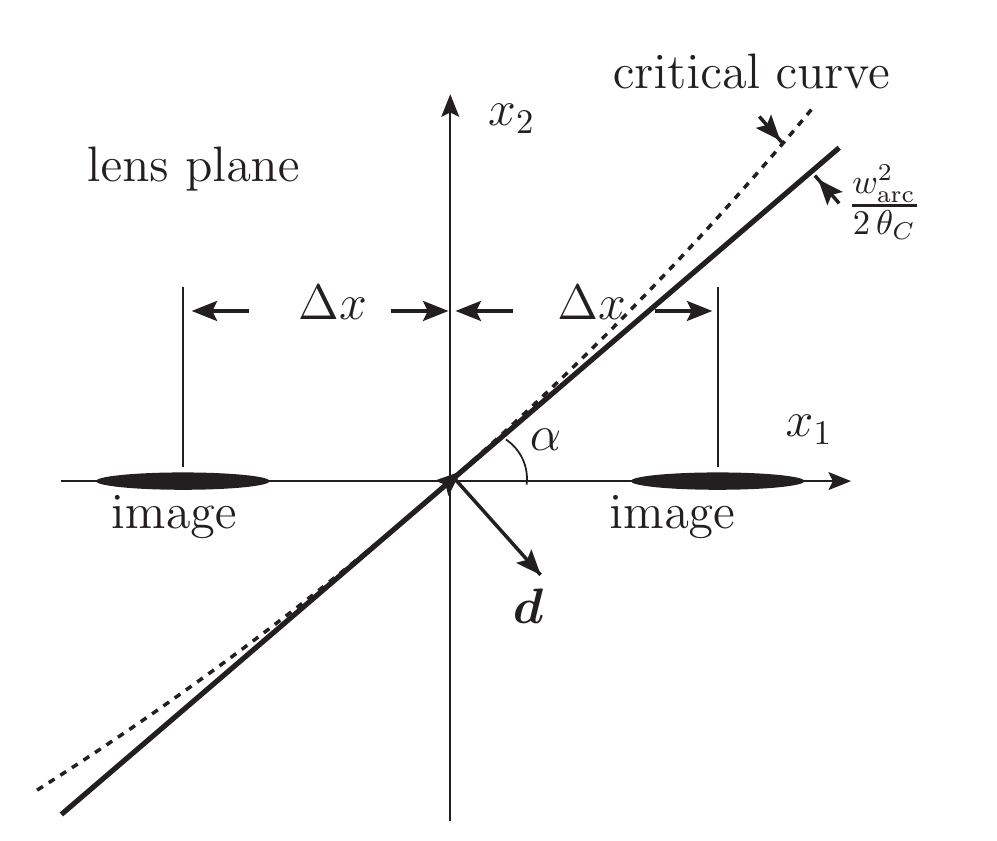}
    \caption{\label{fig:fold} Symmetric image positions in the fold catastrophe of a smooth lens. The critical curve (solid) can be locally approximated as a straight line. The degenerate direction is along the $x_1$ axis. To emphasize the stretching of the images, their sizes have been exaggerated. The dashed line illustrates a critical curve with a small curvature.}
  \end{center}
\end{figure}

Critical curves are contours on the image plane where the magnification diverges, i.e., $\det\bfA(\bfx) = 0$. It is convenient to choose a coordinate system with the origin  on a point of a critical curve around which we wish to examine the lens map, with its first axis ($\hat{\bfx}_1$) along the direction corresponding to the zero eigenvalue of the Jacobian matrix (see \reffig{fold}). If the lens mass is smoothly distributed near the origin (i.e., for small $\bfx$), we can approximate the Jacobian matrix $\bfA(\bfx)$ as~\citep{schneider1992gravitational}
\ba
\label{eq:jacfold}
\bfA(\bfx) =
\left[\begin{array}{cc}
\bfx\cdot \bfd\qquad & 0 \\
0 & 2\,(1 - \kappa_0)\\
\end{array}\right],
\ea
which is characterized by a local convergence $\kappa_0$, and an eigenvalue gradient vector $\bfd = (d\sin{\alpha}, -d\cos{\alpha})$. Locally, the critical curve is the straight line $\bfd\cdot\bfx = 0$, which is inclined at an angle $\alpha$ relative to the $\hat{\bfx}_1$ axis, the direction of elongation of a background galaxy image. The convergence $\kappa_0$ is typically of order unity, and the angular scale $1/d = 1/|\bfd|$ is $\sim$ tens of arcseconds.

Consider an image-plane trajectory $\bfx(\lambda)$ that intersects the critical curve and is parallel to the degenerate direction of the lens map: for the system in \refeq{jacfold}, $\bfx(\lambda) = \lambda \, \hat{\bfx}_1$. The source location $\bfy(0)$ corresponding to $\lambda = 0$ lies on the lensing caustic. We solve for the source--plane trajectory using the definition of the Jacobian matrix as follows:
\begin{align}
  \bfy(\lambda) - \bfy(0) & = \int_0^\lambda d\lambda^\prime \, \bfA(\bfx(\lambda^\prime)) \cdot \frac{d\bfx(\lambda^\prime)}{d\lambda^\prime}. 
\end{align}
Substituting the form of $\bfx(\lambda)$, and the Jacobian $\bfA(\bfx)$ from \refeq{jacfold}, we see that $\bfy$ is locally a quadratic function of $\lambda$. Given a displacement $\Delta y$ toward the inside of the caustic, the solution for $\lambda$ tells us that there are two images separated by
\begin{align}
  2\Delta x 
& = 2 \left( \frac{\Delta y}{|d\,\sin\alpha|} \right)^{1/2} \notag \\
  & = 0.7 \arcsec \times \left( \frac{\Delta y\, D_{\rm S}}{10 \, {\rm pc}} \, \frac{1 \, {\rm Gpc}}{D_{\rm S}} \, \frac{{\rm arcmin}^{-1}}{|d\,\sin\alpha|} \right)^{1/2}.
\end{align}
Both images have the same magnification factor
\begin{align}
|\mu| & = \left[\, 2\,\Delta x\,|(1-\kappa_0)\,d\,\sin\alpha|\,  \right]^{-1},
\end{align}
and are symmetrically arranged about the critical curve along the direction $\hat{\bfx}_1$, as illustrated in Figure \ref{fig:fold}. Each image is itself elongated by the large magnification, although this is unresolvable for stars. 

 When several stars are located near a caustic, each one has a pair of images aligned along the common degenerate direction. The image separations vary, but the critical curve bisects all segments joining each pair. This can be exploited as a model-independent test of the smoothness of the lens model. While the critical curve is not directly observable, it is traced by the image positions of lensed stars.

There are two effects that can potentially change the simple symmetry of the fold: microlensing by intracluster stars, and small-scale structure in the dark matter distribution that perturbs the smoothness of the fold. We will examine microlensing later in \refsec{microlensing} and show that it is unimportant. The natural radius of curvature of the critical line is the Einstein scale of the cluster $\theta_C \simeq 20''$. Across the typical width of a giant arc, $w_{\rm arc} \simeq 0.5''$, this curvature causes a departure from a straight line of only $\sim w^2_{\rm arc}/(2\,\theta_C) \simeq 10\,{\rm mas}$ (see \reffig{fold}), about one third of the pixel size of {\em HST}. Larger curvatures can arise from the lensing influence of individual cluster galaxies if they happen to lie close in projection. We will show that dark matter subhalos produce larger curvatures and more complex distortions.

\subsection{Lensing with subhalos}
\label{sec:subhalolens}

\begin{figure}[t]
  \begin{center}
    \includegraphics[scale=0.55]{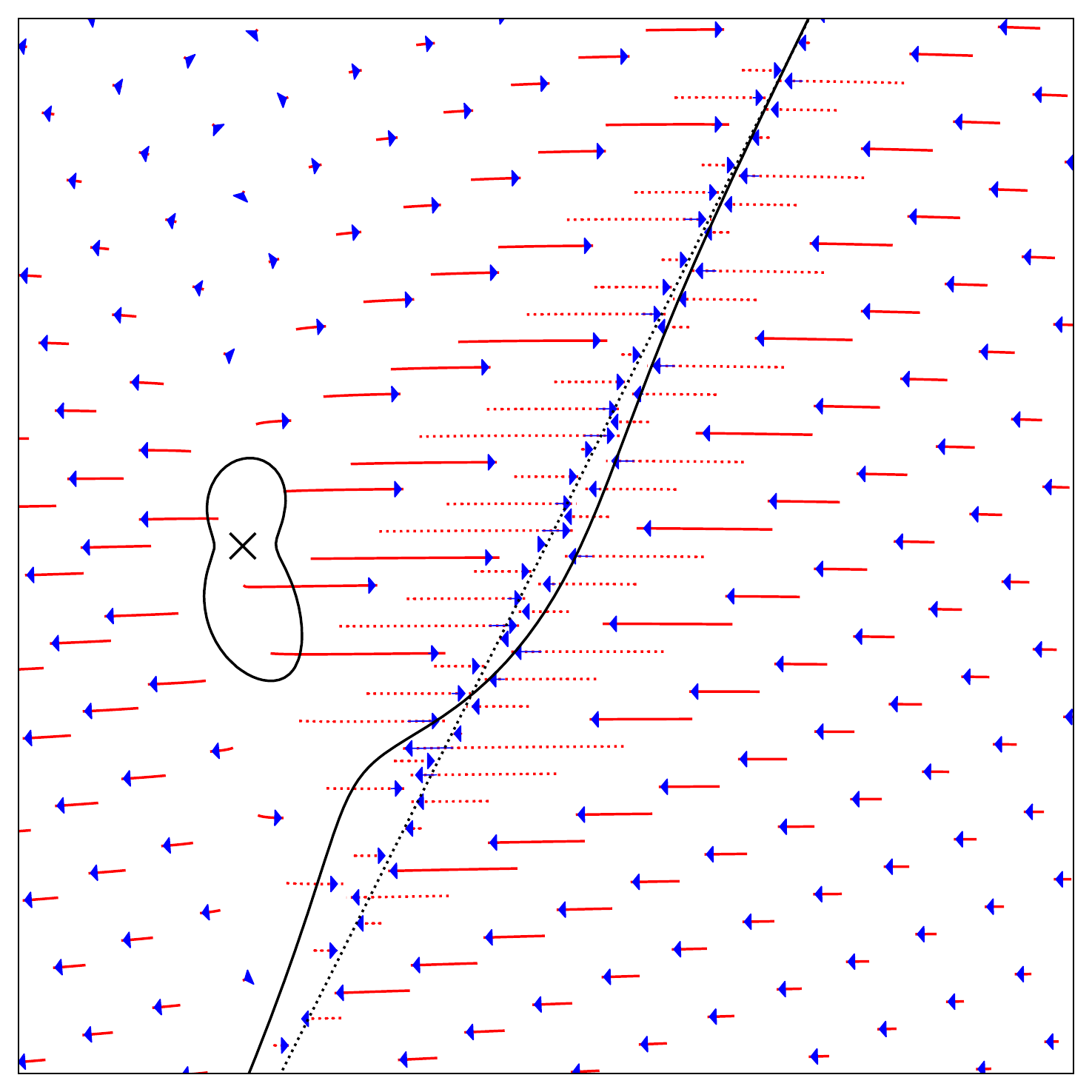}
    \includegraphics[scale=0.55]{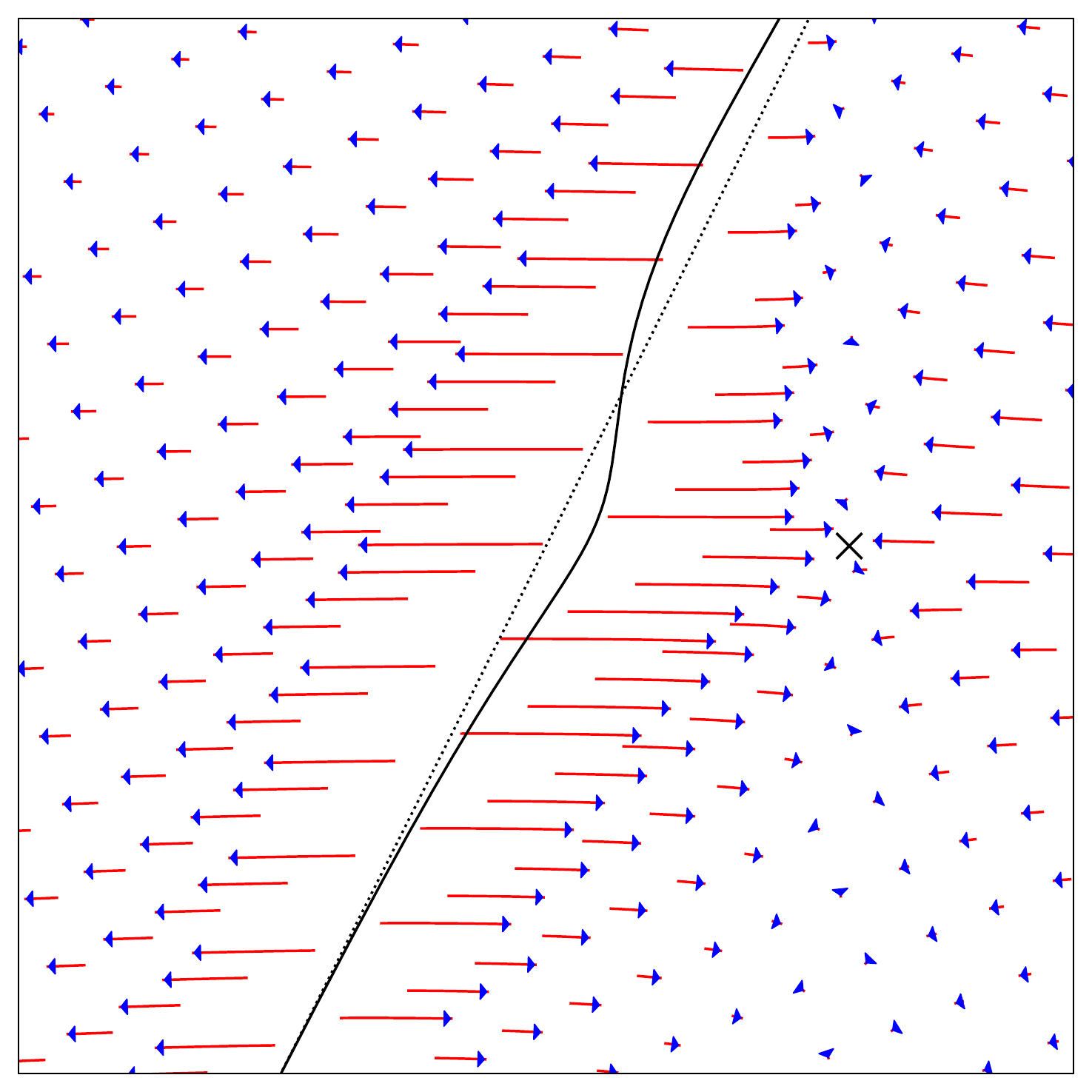}
    \caption{\label{fig:subhalo_demo} Perturbations to image locations and the critical curve due to a single subhalo centered on the black cross. Dotted and solid lines are the lensing critical curves without and with a subhalo, respectively. Arrows indicate image displacements as the subhalo mass is increased. Dashed segments indicate images that merge with their counter-image and disappear before the subhalo mass reaches its final value. The fold parameters used are typical of the critical curves of strong lensing galaxy clusters. The subhalo is modeled as a truncated NFW sphere with bound mass $m=2.5\times 10^7\,M_\odot$, scale radius $r_s = 300\,{\rm pc}$, and concentration parameter $c_{200} = 24$ (see \refsec{subhalo_internal}). {\it Upper panel:} Subhalo located to the outside of the critical curve ($\bfd\cdot\bfx >0$). Images very close to the critical curve shift toward it, and some eventually merge with their counter-images and disappear. {\it Lower panel:} The same subhalo located to the inside of the critical curve ($\bfd\cdot\bfx <0$). Images very close to the critical curve shift apart from it, and new pairs of images emerge (not shown in this figure; they fill the vacated space on both sides of the critical curve). Here we have always assumed $\kappa_0 <1$, which is the case for many caustic-crossing arcs. The case of over-focusing $\kappa_0 > 1$ can be similarly studied.}
  \end{center}
\end{figure}

We now consider the effect of cluster subhalos in the projected vicinity of a critical curve.
Figure \ref{fig:subhalo_demo} illustrates how a single subhalo perturbs the critical curve. The perturbed critical curve has a characteristic wiggle over an angular region comparable to the separation between the subhalo center and the unperturbed critical curve, with an amplitude that grows with the subhalo mass and decreases with the separation. Owing to the quadrupolar nature of the subhalo shear, the perturbation consists of two ``bulges'' facing opposite directions, rather than a single ``bulge''. A new closed branch of the critical curve may arise around the subhalo center if its perturbation is sufficiently strong (as is the case in \reffig{subhalo_demo}). 

Dark matter subhalos are generally difficult to detect through lensing because their surface density is highly subcritical. However, near the critical line of a lensing cluster, their effect is greatly enhanced. This can be seen from the following argument, first presented in~\cite{1986ApJ...306....2K}.

In the absence of the perturber, the image position $\bfx$ satisfies the lens equation
\ba
\label{eq:lens_eq_bkg}
\bfx - \bfy - \bfalp_{\rm B}(\bfx) = 0,
\ea
where $\bfalp_{\rm B}(\bfx)$ is the deflection angle of the smooth lens model with no subhalos. Adding a subhalo shifts the image to a new position $\bfx' = \bfx + \Delta\,\bfx$, which satisfies the new lens equation
\ba
\label{eq:lens_eq_ptb}
\bfx + \Delta\,\bfx - \bfy - \bfalp_{\rm B}(\bfx + \Delta \bfx) - \bfalp_{\rm sh}(\bfx') = 0 ~.
\ea
Expanding to first order the smooth deflection as $\bfalp_{\rm B}(\bfx + \Delta \bfx) \approx \bfalp_{\rm B}(\bfx) + \Delta\bfx\cdot\boldsymbol{\nabla}\, \bfalp_{\rm B}(\bfx)$, the shift of the image position in \refeq{lens_eq_ptb} is
\ba
\label{eq:atmshift}
\Delta \bfx \approx \left[ \bfA_{\rm B}(\bfx) \right]^{-1} \cdot \bfalp_{\rm sh}(\bfx').
\ea
The smooth Jacobian matrix $\bfA_{\rm B}(\bfx)$ is nearly degenerate close to a fold, as in \refeq{jacfold}. Its inverse has a large eigenvalue $\mu(\bfx)/[2\,(1-\kappa_0)]$ along the degenerate direction, where $\mu(\bfx)$ is the magnification factor in the fold model. The image displacement $\Delta\bfx$ is of the same order as the deflection $\bfalp_{\rm sh}(\bfx')$ along the non-degenerate direction, but is larger by a factor of $\mu$ along the degenerate direction. The astrometric perturbation of subhalos is therefore dramatically amplified near the critical curve, making highly magnified image pairs sensitive probes of substructure.

As seen in Figure \ref{fig:subhalo_demo}, the image displacements caused by subhalos are predominantly along the degenerate direction. In line with \refeq{atmshift}, images that are close to the smooth critical curve exhibit the largest displacements under the subhalo influence, while distant images are less affected. These displacements break the symmetry of image pairs with respect to the smooth critical curve. In the vicinity of the critical curve, a pair of images corresponding to the same source are displaced in opposite directions but by asymmetric amounts, causing their midpoint to shift.

If the subhalo has a closed branch of the critical curve around its center, a source located within the corresponding closed caustic has two image pairs (instead of one) which are aligned along the common degenerate direction. If detected, this would immediately exclude a smooth lens model. However, because of the small area within closed caustics of subhalos, we expect these cases to be rare. 

\subsection{Microlensing by intracluster stars}
\label{sec:microlensing}

The projected surface density of intracluster stars near critical curves, typically $\sim 50\,{\rm kpc}$ from the cluster center, usually contributes a lensing convergence $\kappa_\star \simeq 10^{-3}$--$10^{-2}$~\citep{1951PASP...63...61Z, 2004ApJ...617..879L, 2005MNRAS.358..949Z}. These stars may have been tidally stripped from cluster galaxies throughout the assembly history of the brightest central galaxy (BCG), or have formed {\it in situ} within tidally ejected intracluster clouds~\citep{Martel:2012ue,Contini:2013wha,Cooper:2014nwa}. Despite these low values of $\kappa_\star$, a similar line of reasoning as that in \refeqs{lens_eq_bkg}{atmshift} shows that the associated optical depth to microlensing is enhanced by the magnification factor of the fold model~\citep[e.g.,][]{Venumadhav:2017ttg}. This is the reason for the substantial flux variations in the individual images of the caustic-crossing star detected in observations of MACS J1149~\citep{2017arXiv170610279K,Diego:2017drh,Oguri:2017ock}.

Microlensing breaks the smooth critical curve into a band of micro-critical curves of width $\sim 2\,\kappa_\star/d$, where images are generally affected by more than one microlens. There are two main effects on the macroimages of background stars i.e., their images in a model without microlenses.

First, each macroimage is broken into a series of aligned microimages along the degenerate direction, with a typical angular spread $\lesssim r_f \simeq  \theta^{1/2}_\star\,\kappa^{1/4}_\star/|d\,\sin\alpha|^{1/2} \simeq 1\,{\rm mas}$, where $\theta_\star$ is the Einstein radius of a microlens~\citep{Venumadhav:2017ttg}. This spread is not resolvable by present telescopes, so each track of micro-images appears as a single macroimage.

Second, the number of microimages and their fluxes fluctuate due to the differential proper motion of the source and lens. The net result is a stoachastic variation in the flux of each macroimage, and a jitter in the position of its centroid. The latter is at the $\sim 1 \, {\rm mas}$ level. This is negligibly small compared to the astrometric signature ($\sim 10$--$100\,{\rm mas}$) of subhalos with masses of $10^{6-8} \, M_\odot$. Thus we expect our astrometric method to be unaffected by microlensing.

Microlensing variations in the flux may sometimes push individual macroimages below the detectability threshold. Indeed, this occurred during the observation of the first caustic crossing star in MACS J1149~\citep{2017arXiv170610279K}. This may complicate the detection and identification of image pairs, but can be corrected through imaging at multiple epochs to average over microlensing variations. In general, our method does not have to rely on the flux to correctly identify image pairs associated with the same source star. Instead, one may use the properties of image pairs laid out in Sec.~\ref{sec:fold}---macro images of the same source star almost always align along the common degenerate direction. Colors are also helpful in identifying image pairs.

\section{Case study: giant arc in Abell 370}
\label{sec:sim}

\begin{figure}[t]
  \begin{center}
    \includegraphics[width=\columnwidth]{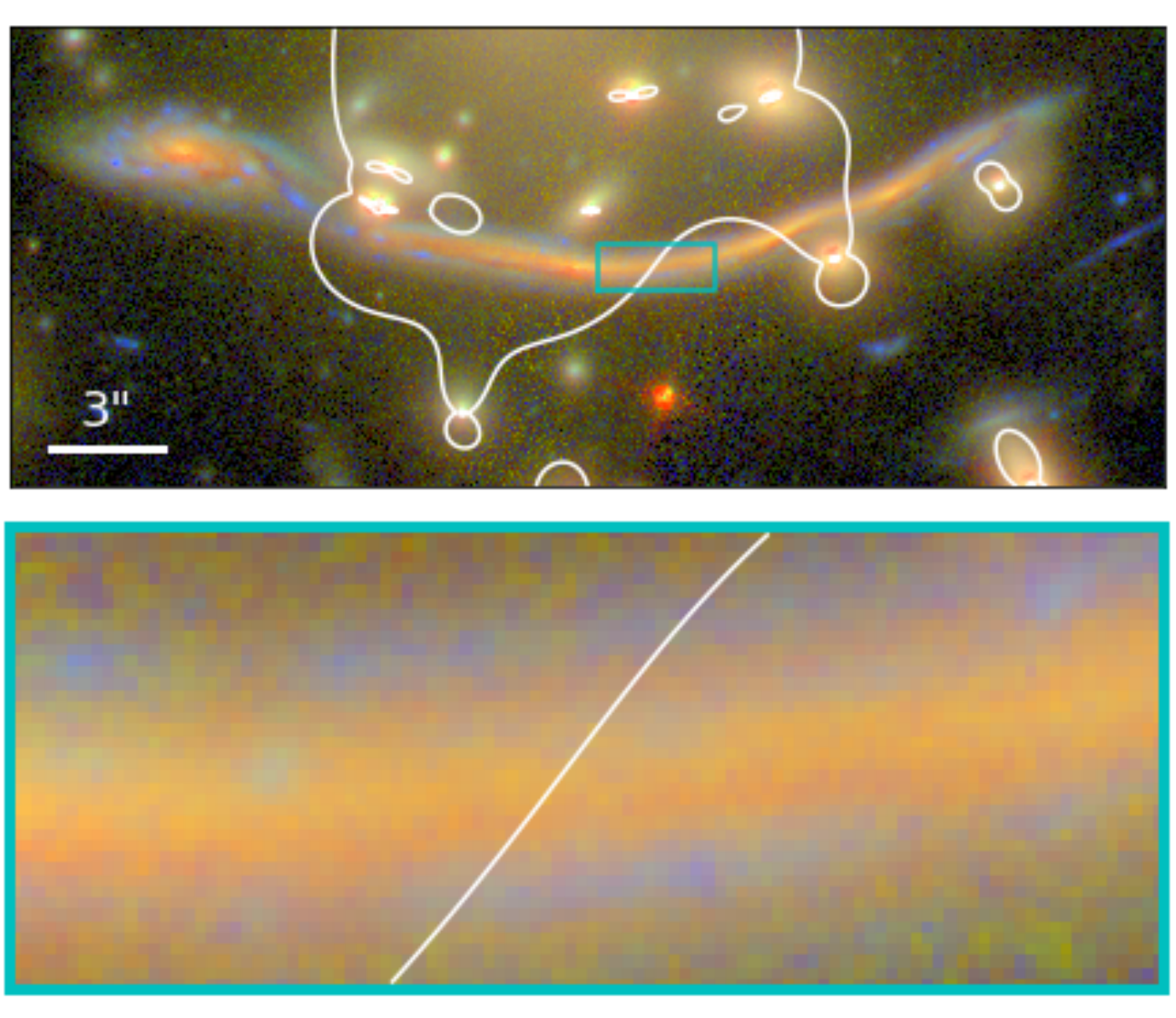}
    \caption{\label{fig:a370} {Upper panel}: False color image of the ``Dragon'', a giant arc in Abell 370 , plotted using the archival {\it HST} data from {\sf f450w}+{\sf f606w} (blue), {\sf f814w} (green), and {\sf f105w}+{\sf f125w} (red) filters. White line is the critical curve at the arc redshift $z_s=0.725$. {Lower panel}: zoomed-in region of size $3\arcsec\times 1.2 \arcsec$ of the most promising crossing location of the critical curve. The lens model is adopted from \citet{2010MNRAS.402L..44R} and the critical curve was calculated using {\sf Lenstool}~\citep{2009MNRAS.395.1319J}.}
  \end{center}
\end{figure}

We illustrate our method via a case study of the massive galaxy cluster Abell 370~\citep{abell1958distribution} at $z_l = 0.375$~\citep{mellier1988photometry}, historically one of the first studied strong lensing systems~\citep{fort1988faint, mellier1991spectroscopy, kneib1993distribution, smail1995hst, bezecourt1998lensed, bezecourt1999lensed}. The cluster is notable for the ``Dragon'', an exceptionally long and luminous arc~\citep{lynds1986giant,soucail1987blue} which is a gravitationally lensed background galaxy at a source redshift of $z_s = 0.725$~\citep{soucail1988giant, lynds1989luminous}. Detailed lens modeling shows that the giant arc consists of five images of the background galaxy joined end-to-end at four intersections with the lensing critical curve~\citep{2010MNRAS.402L..44R, 2017MNRAS.469.3946L} (see upper panel of \reffig{a370}).

In this section, we present lensing simulations of one of these intersections in the ``Dragon'', including a population of subhalos in its projected vicinity, and we discuss observational prospects for the method discussed in \refsec{subhalolens}. Our simulations contain the following ingredients: (1) a smooth fold in a region near our chosen intersection of the critical curve and the giant arc; (2) random realizations of subhalos causing substantial lensing perturbations in this region; (3) a realistic, randomly generated stellar population in the corresponding region of the source galaxy. We then calculate the positions and fluxes of the star images.

We do not include microlensing by intracluster stars in our simulations. As previously argued, this does not affect the image positions, but causes additional flux variations. A full-scale simulation of the entire intersection including microlensing was not computationally feasible with our method.

\subsection{Cluster mass profile and the fold model}
\label{sec:host}

The mass density profile of galaxy clusters in numerical simulations is on average well described by the Navarro-Frenk-White (NFW) model, which is parameterized by a virial mass $M_{200}$, scale radius $R_{\rm s}$, and concentration parameter $C_{200}$~\citep{1996ApJ...462..563N} (see the Appendix for precise definitions). Typical values for dynamically relaxed clusters are $M_{200} \simeq 2\times10^{15}\,M_\odot$ and $C_{200} \simeq 7$--$8$~\citep{umetsu2011cluster}. 

The mass modeling for Abell 370 is complicated by the dynamically unrelaxed nature of the cluster, which has two components of comparable mass that are undergoing a major merger nearly along the line of sight~\citep{2010MNRAS.402L..44R}. The two components are centered near the two brightest central galaxies (BCGs)---a primary one near the ``Dragon'', and a secondary one about $200\,{\rm kpc}$ away in projection.

We model the total mass profile by the superposition of two equal NFW halos, each with virial mass $M_{200} = 1.6\times 10^{15}\,M_\odot$, concentration parameter $C_{200} = 7$, scale radius $R_s = 310\,{\rm kpc}$, and virial radius $R_{200} = C_{200}\,R_s = 2.1\,{\rm Mpc}$, respectively. The resulting total enclosed mass reasonably reproduces the measured value of $M(< 250\,{\rm kpc}) = 3.8\times 10^{14}\,M_\odot$ centered at the midpoint of the two BCGs~\citep{2010MNRAS.402L..44R}. The concentration parameters found by \cite{umetsu2011cluster} are significantly higher than typical values in N-body simulations for similar halos. This is not uncommon among other strong lensing clusters~\citep[e.g., Abell 2667;][]{covone2006vimos}, and is likely due to a selection bias for concentrated systems that have larger cross sections for strong lensing.

The ``Dragon'' intersects the critical curve at four locations. We focus on a small region of size $1'' \times 1''$ centered on one of these intersections, shown in the lower panel of \reffig{a370}. The impact parameter from the primary BCG is $\simeq 10''$, or $B = 0.17\,R_s = 53\,{\rm kpc}$ in proper units. The local fold corresponding to the smooth lens model has parameters $\kappa_0 = 0.69$, $|\bfd|=2.7\,{\rm arcmin}^{-1}$, and $\alpha = 45^{\circ}$ [see \refeq{jacfold}]. We select this fold location because of the higher magnification factors compared to the other three locations.

\subsection{Subhalo abundance}
\label{sec:subhalo_abundance}

Generating random realizations of subhalos requires a model of their abundance and spatial distribution in cluster halos. \cite{Han:2015pua} derive such a model from N-body simulations with moderate subhalo-to-host mass ratios $10^{-7} < m/M_{200} < 10^{-3}$. However, subhalo masses of our interest ($m \lesssim 10^8\,M_\odot$ as we will see later) are under-resolved even in state-of-the-art simulations of cluster-size halos. Hence, we need to extrapolate their model to lower subhalo masses.

Based on the semi-analytic model of \cite{Han:2015pua}, we use the following procedure to generate subhalos. We first generate the initial subhalo mass $m_{\rm acc}$ when the subhalo is incorporated into the host halo. We assume that the {\it unevolved} specific mass function, $dn(m_{\rm acc},R)/d\log\, m_{\rm acc}$, spatially traces the host halo mass density profile $\rho(R)$,
\ba 
\label{eq:dndlnmacc}
\frac{dn(m_{\rm acc},R)}{d\log\,m_{\rm acc}} = A_{\rm acc}\,\frac{\rho(R)}{m_0}\,\left( \frac{m_{\rm acc}}{m_0} \right)^{-\alpha},
\ea
where $m_0 = 10^{10}\,h^{-1}\,M_\odot$. We adopt commonly used values for the power-law index $\alpha=0.9$~\citep{mo2010galaxy} and normalization constant $A_{\rm acc} = 0.08$~\citep{Han:2015pua}. 

\begin{figure*}[t]
  \begin{center}
    \includegraphics[scale=0.7]{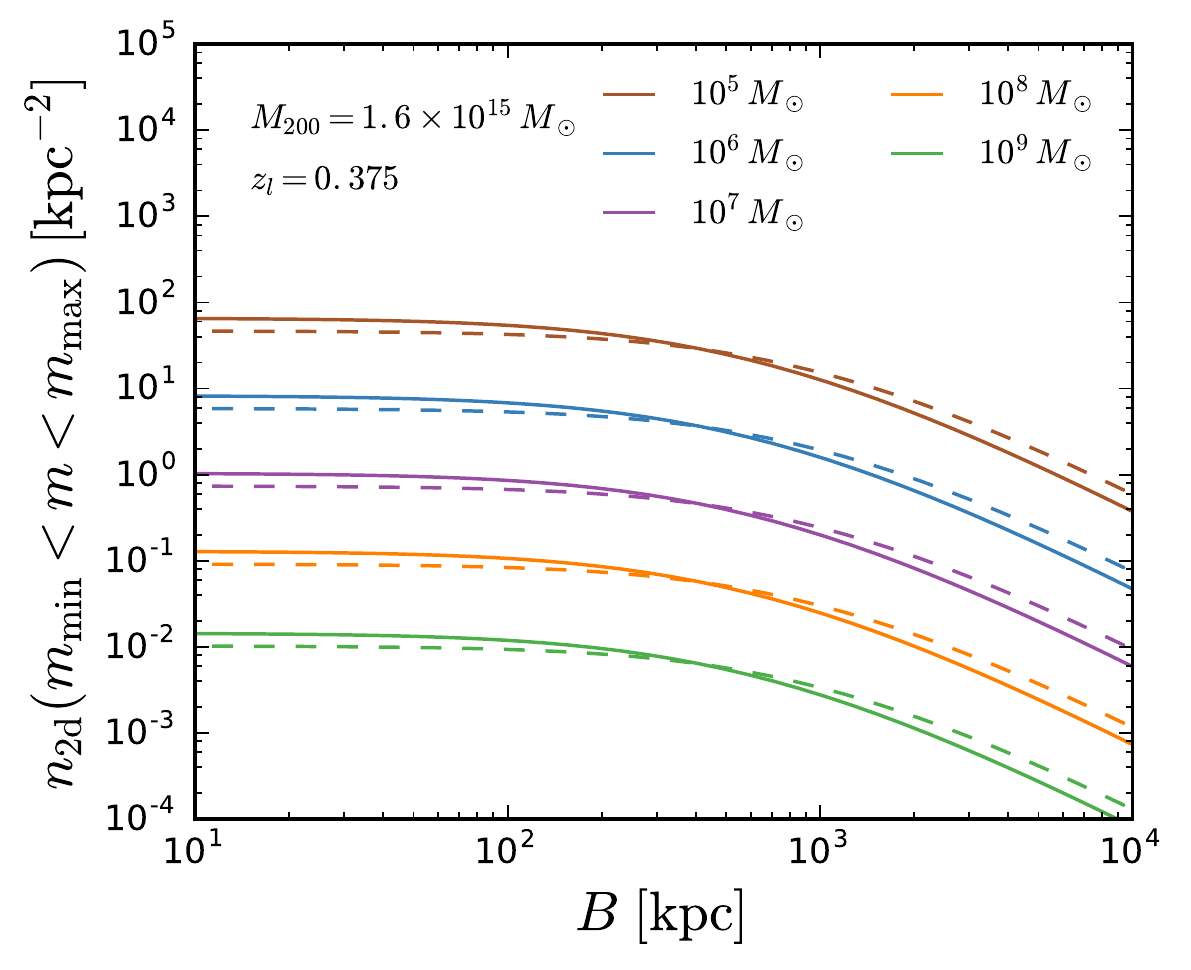}
    \includegraphics[scale=0.7]{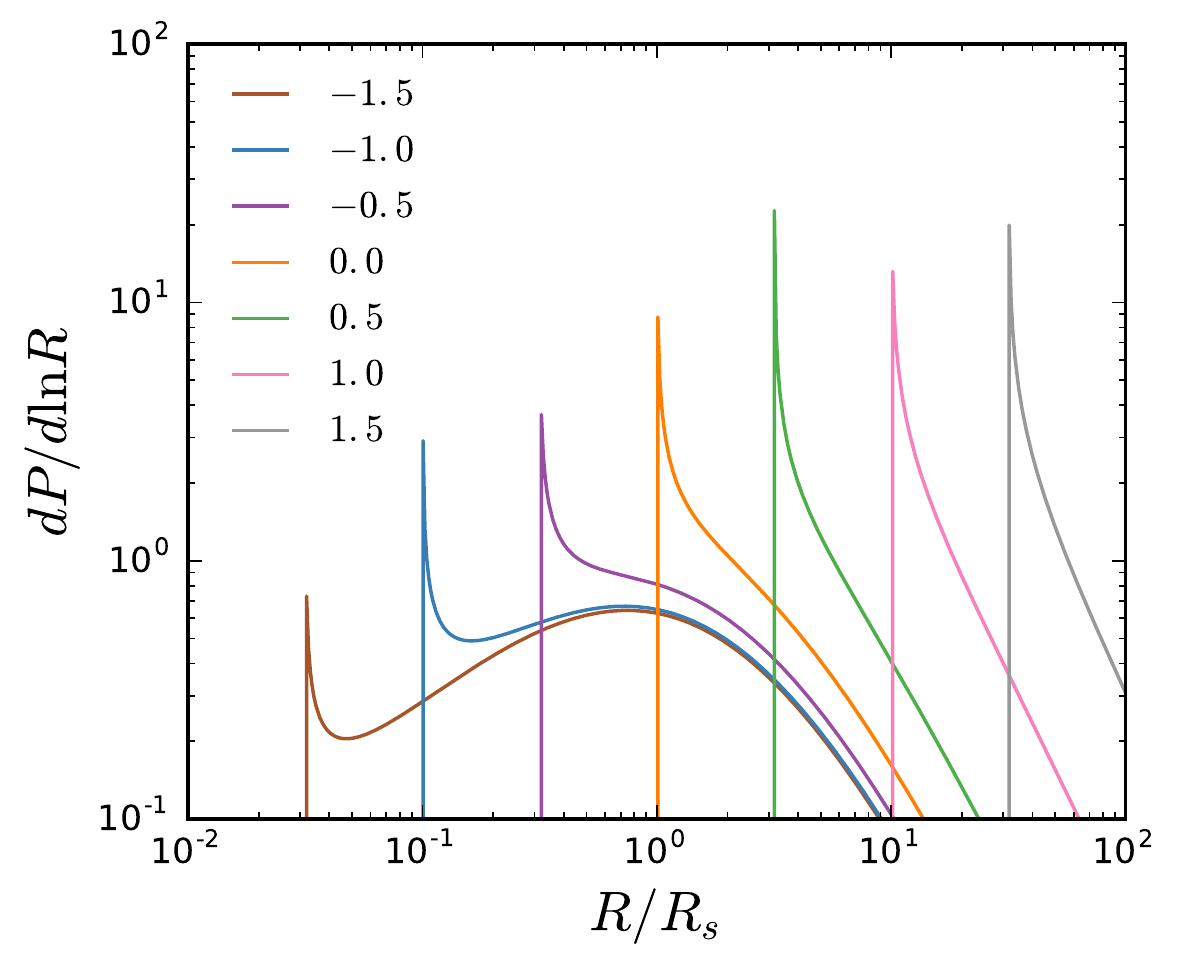}
    \caption{\label{fig:shabun} {\it Left panel}: Projected number density of subhalos with $m_{\rm min} < m < m_{\rm max} = 10^{10}\,M_\odot$ versus the impact parameter $B$ of the sight line for galaxy cluster Abell 370. Colors correspond to various minimum subhalo masses $m_{\rm min}$. Solid lines are computed for the fiducial halo parameters mentioned in the text. Dashed lines are for $C_{200} = 3.5$. {\it Right panel}: Probability distribution of finding a subhalo along the line of sight having a halocentric distance $R$. Curves correspond to various values of $\log(B/R_s)$ as indicated in the legend.}
  \end{center}
\end{figure*}

Owing to tidal stripping, the bound mass $m$ of a subhalo decreases with time from its initial value $m_{\rm acc}$. In simulations, the ratio $m/m_{\rm acc}$ is highly stochastic, since tidal stripping depends on the orbital history of the subhalo. We present our procedure for generating the distribution of $m/m_{\rm acc}$ in detail in \refapp{shabund}. In brief, and following \cite{Han:2015pua}, we draw the ratio $m/m_{\rm acc}$ from the log-normal distribution of \refeq{mfrommacc}, which depends on the halocentric radius $R$ and includes a probability for complete tidal disruption that is set to 56\%.

The left panel of \reffig{shabun} shows the projected subhalo abundance as a function of the projected distance $B$ to the primary BCG. The surface number density increases with decreasing distance $B$ down to the scale radius $R_s$ of the host, and plateaus thereafter. Note that the caustic crossing location on the giant arc has an impact parameter $B = 53\,{\rm kpc}$ to the nearby BCG, which is much less than the scale radius $R_s$. Since the projected separation between the two component halos is also less than $R_s=310\,{\rm kpc}$, the subhalo contributions from both components are nearly equal for equal component masses. Given the mass-modeling uncertainty for the non-relaxed cluster, we generate subhalos within the NFW halo associated with the nearby BCG using \refeq{dndlnmacc}, and simply multiply the abundance by a factor of two. Including both components, we find on average $\simeq 1$ subhalo more massive than $10^7\,M_\odot$ within a $0.2'' \times 0.2''$ patch. Subhalo lensing is therefore not rare, and we have checked that the result does not substantially change even if we set the concentration parameter to its cosmic mean of $C_{200} = 3.5$ for cluster-sized halos at low redshifts~\citep{ludlow2016mass}. 

The right panel of \reffig{shabun} shows that even for low impact parameters, the subhalos in the projected vicinity of the line of sight are most likely to have halocentric distance $R \sim R_s$. Therefore, although the impact parameter is typically $B < R_s$, most detectable subhalos that our method is sensitive to are likely far in three-dimensional distance from the central region near the BCG.

For the lensing simulations, we generate subhalos within a cylinder of radius $\mathcal{R} = 1''$ and line-of-sight depth $100\,R_s$ centered on the simulated $1'' \times 1''$ region. We conserve the mass within the cylinder by removing an appropriate uniform surface-mass density. We focus on subhalo masses $m \in [10^4,\,10^9] M_\odot$, since more massive subhalos are rare, while subhalos with lower masses do not have a resolvable astrometric signature.

\subsection{Internal structure of subhalos}
\label{sec:subhalo_internal}

The lensing effects of subhalos depend on their detailed mass profiles. Our method probes subhalos with mass lower than that of the host cluster halo by factors of $\sim 10^{7-9}$. Such a large dynamic range has never been achieved even in the highest resolution dark-matter only simulations run so far~\citep{2008Natur.454..735D}. Recent work has shown that simulations are prone to artificial numerical relaxation effects at the edge of their mass resolution~\citep{2018MNRAS.474.3043V}; this precludes a direct use of numerically obtained subhalo mass profiles near the resolution limit. We will extrapolate simple fits to the profiles of more massive subhalos and neglect complications such as triaxiality. 

We parameterize subhalo density profiles as smoothly truncated NFW (TNFW) profiles~\citep{Baltz:2007vq, Cyr-Racine:2015jwa}, with the functional form
\ba
\label{eq:shrho}
\rho_{\rm sh}(r) = \frac{m_{200}}{4\pi r^3_s\,f(c_{200})}\,\frac{1}{(r/r_s)(1 + r/r_s)^2}\,\frac{1}{1+r^2/r^2_t}\, , ~
\ea
where $r$ is the distance to the subhalo center, and the finite total mass $m$ differs from $m_{200}$. The parameters $m_{200}$, $r_s$ and $c_{200}$ are the usual parameters of (untruncated) NFW halos. The instantaneous tidal radius $r_t$ is defined in terms of the halocentric distance within the host halo, $R$, as~\citep{binney1987galactic}
\ba
\label{eq:instidal}
r_t = \left( \frac{m}{M_<(R)} \right)^{\frac13}\,\left( 3 - \frac{d\ln M_<(R)}{d\ln R} \right)^{-\frac13}\,R,
\ea
where $M_<(R)$ is the mass enclosed within $R$. At fixed $m$, this always underestimates the true compactness of the subhalo, which is generally truncated at a radius smaller than $r_t$.

We use a concentration parameter that increases toward the inner part of the host halo according to the fit~\citep{2001MNRAS.321..559B, 2007ApJ...667..859D, 2015PhRvD..92l3508B, 2017MNRAS.466.4974M}
\begin{multline}
c_{200}(m, R, z) = \\
\bar{C}_{200}(m, z) \left[ 1 + \frac{1}{15}\left( \frac{(1.5\,R_{200})^2}{R^2 + (0.1\,R_{200})^2} \right)^{\frac12} \right], \label{eq:shc200}
\end{multline}
where $\bar{C}_{200}(m, z)$ is the concentration for field halos~\citep{ludlow2016mass}. We collect some analytical formulae for lensing by TNFW density profiles in \refapp{shlen}.

\subsection{Source stellar population}
\label{sec:stellar}

The final ingredient in our simulations is the magnitude distribution of stars in the vicinity of the caustic. We model the source stellar population based on archival {\it HST} images of Abell 370 in seven filters: {\sf f435w}, {\sf f606w}, {\sf f814w}, {\sf f105w}, {\sf f125w}, {\sf f140w} and {\sf f160w}.

We model the data within a $1'' \times 1''$ square patch that is centered on the midpoint of the segment of the lensing critical curve shown in the zoomed-in panel of \reffig{a370}, and has its first coordinate axis aligned with the direction of arc elongation. Based on the surface brightness and color inferred from the multi-band {\it HST} data, we divide the patch into five regions, and separately model the stellar population in each region using the following procedure.

\begin{figure*}[t]
  \begin{center}
  	\includegraphics[width=0.975\columnwidth]{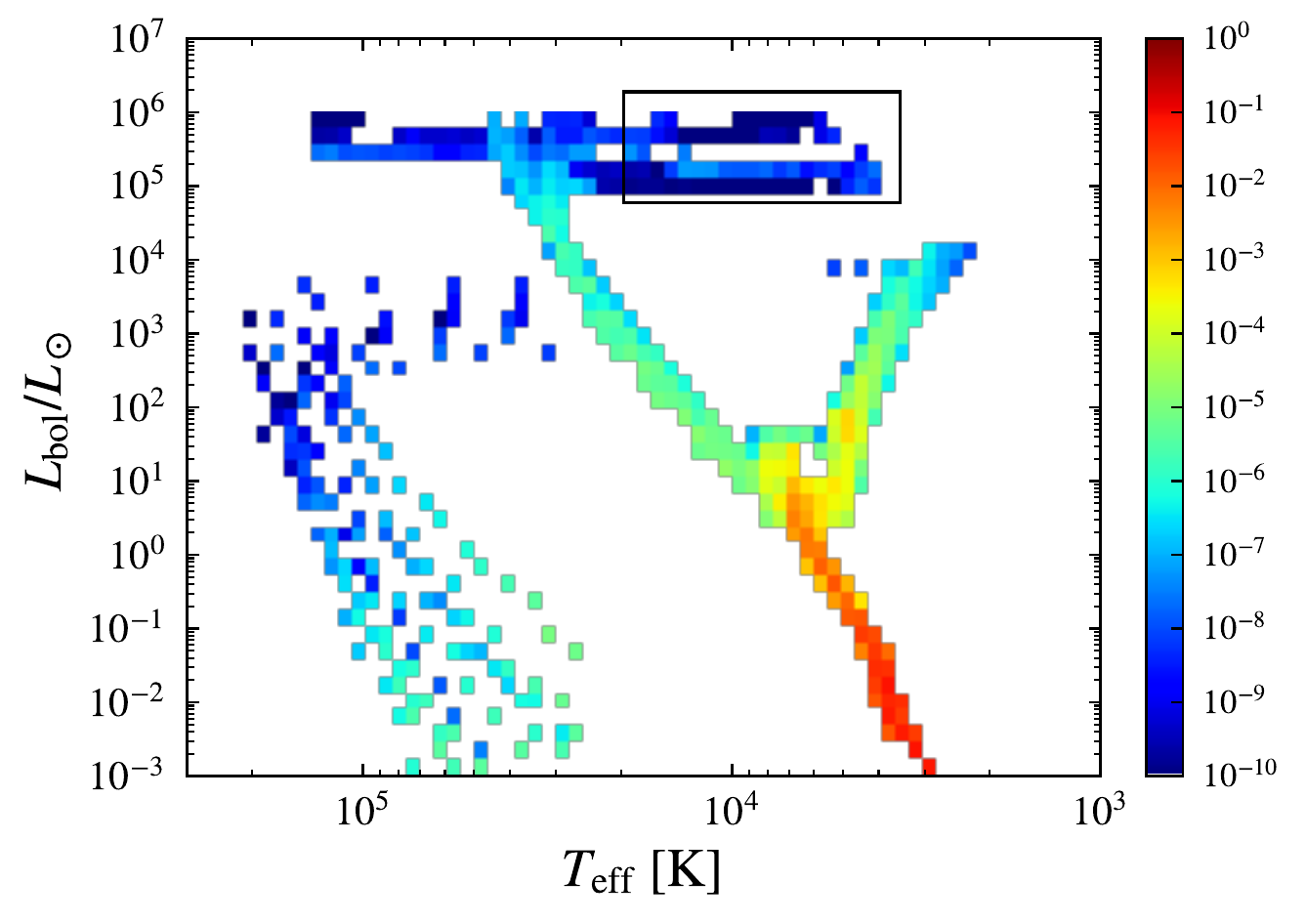}
    \includegraphics[width=1.025\columnwidth]{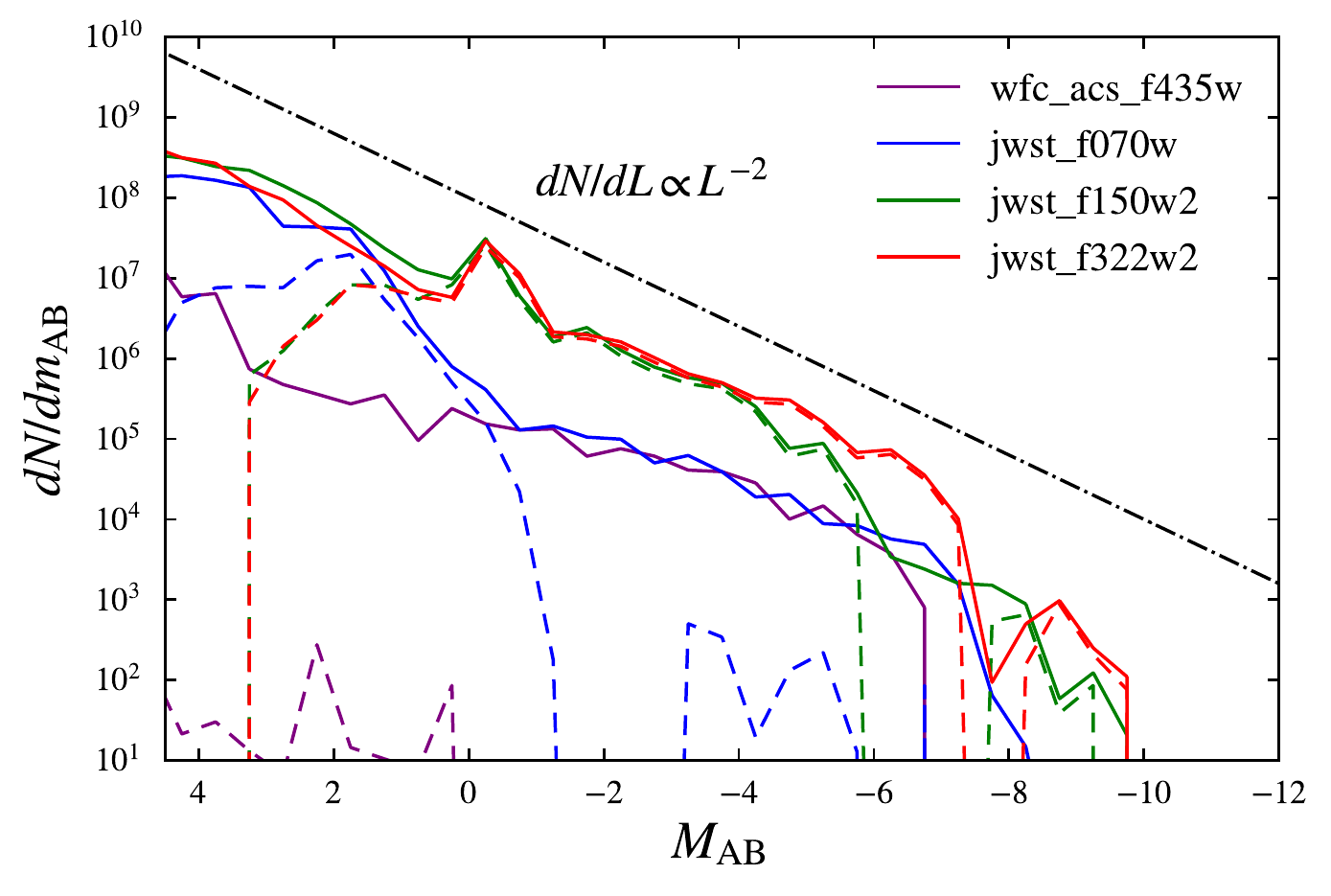}
    \caption{\label{fig:dNdlogL} {\it Left panel}: Joint distribution of bolometric luminosity $L_{\rm bol}$ and surface temperature $T_{\rm eff}$ for the source stellar population at our chosen caustic crossing location in the strong lensing system of Abell 370. Bins in $L_{\rm bol}$ and $T_{\rm eff}$ are chosen to be log-uniform. The color-coded value is proportional to the number of stars in each $(L_{\rm bol}, T_{\rm eff})$ bin. The black rectangle marks the region of supergiants in the diagram where stars can be sufficiently bright at optical/near-IR wavelengths to be individually detectable. {\it Right panel}: Distribution of the absolute magnitude $M_{\rm AB}$ in a number of filters (dust attenuated). Two cases are computed for comparison: (1) all stars included (solid); (2) only red stars having $T_{\rm eff}< 5200\,$K (dashed). All histograms are normalized to a fixed but arbitrary total number of stars. For visual guide, the black dash-dotted line shows the power-law $dN/dL \propto L^{-2}$.}
  \end{center}
\end{figure*}

We consider six values of stellar metallicity: $\log(Z/Z_\odot) = 0.2,\,0.0,\,-0.3,\, -0.5,\, -1.0,\, -1.5$. For each metallicity, we use the population synthesis code $\mathsf{FSPS}$~\citep{conroy2009propagation, conroy2010propagation} to construct simple stellar populations (SSPs) in 14 age bins: $t_{\rm age}=300\,{\rm kyr}$, $500\,{\rm kyr}$, $1\,{\rm Myr}$, $2\,{\rm Myr}$, $5\,{\rm Myr}$, $10\,{\rm Myr}$, $25\,{\rm Myr}$, $50\,{\rm Myr}$, $100\,{\rm Myr}$, $250\,{\rm Myr}$, $500\,{\rm Myr}$, $1\,{\rm Gyr}$, $3\,{\rm Gyr}$ and $7\,{\rm Gyr}$ (the age of the Universe at $z_s = 0.725$ is about $7\,{\rm Gyr}$). We then find the restricted linear combination of age bins (with non-negative coefficients) that best reproduces the measured fluxes in the seven {\it HST} filters. When constructing the SSP templates, we correct for dust attenuation in the source galaxy using the two-component power-law prescription of \cite{charlot2000simple}. We also allow for an additional component to account for the contamination from a nearby cluster member galaxy (only $\simeq 3''$ away in projection), and from the intracluster light surrounding the primary BCG. We restrict this component to have the same color as that of the cluster member galaxies. Finally, we choose the value of metallicity $\log(Z/Z_\odot)$ with the smallest best-fit residual. Although the number of linear components invoked exceeds the number of {\it HST} filters used, our approach does not have an over-fitting problem. We found that the physical requirement that the coefficients of the linear components be non-negative is sufficiently restrictive so as to prevent over-fitting.

\subsection{Simulating caustic crossing stars}
\label{sec:ccssim}

Our fits favor $-0.5 \lesssim \log(Z/Z_\odot) \lesssim -0.3$ in regions of high surface brightness, which is consistent with the mean stellar metallicity found across the entire source galaxy in \cite{patricio2018kinematics}. We identify SSP components with ages in two widely separated ranges, $t_{\rm age}=2$--$10\,{\rm Myr}$ and $t_{\rm age}=1$--$7\,{\rm Gyr}$, with the former suggesting recent star formation. This is in line with the appearance of the galaxy in the reconstructed source-plane images based on multi-band {\it HST} images, with a red core, and blue star-forming clumps in the outskirts. 

In the left panel of \reffig{dNdlogL}, we show the Hertzsprung-Russell diagram for the best-fit stellar population. In general, stars with very high bolometric luminosities $L_{\rm bol} \gtrsim 10^5\,L_\odot$ are promising candidates for individual detection as caustic crossing stars. Specific to $z_s = 0.725$, the brightest stars in the near-IR bands ($1$--$4\,\mu$m) are typically red supergiants of spectral type K. The brightest stars in optical bands ($0.4$--$0.8\,\mu$m) are white and blue supergiants of spectral type A or B and with surface temperatures $T_{\rm eff}=7500$--$20000\,$K. Main-sequence stars of spectral type ranging from B to O can have higher temperatures $T_{\rm eff} \gtrsim 20000\,$K, but their bolometric luminosities are only comparable to those of the supergiants. This is because their typical stellar radii $\sim 10\,R_\odot$ are much smaller than those of the supergiants $\sim 10^2$--$10^3\,R_\odot$. Since most of their energy output is in the UV ($\lambda_{\rm peak} \lesssim 0.15\,\mu$m in the rest frame), hot main sequence stars are not ideal candidates for detection in optical/IR bands~\citep{2017arXiv170610279K}.

The right panel of \reffig{dNdlogL} shows distributions of the absolute magnitude in a few selected filters. The distributions have power-law tails at the bright end, $dN/dL \sim L^{-2}$. This slope compares well to the values measured in actively star-forming systems, such as 30 Doradus~\citep{2017arXiv170610279K}. Caustic-crossing arcs that lack recent star formation are unlikely to host a substantial number of supergiants. Thus, giant arcs that host star-forming structures are more promising targets for detecting caustic crossing stars.

To predict the magnitudes of observable highly magnified stars, we randomly draw stars from the best-fit stellar population and distribute them uniformly on the source plane (in reality young stars cluster; we recall that image pairs of bright clumps such as open clusters or HII regions can be equally used to trace the critical curve). We read off the stellar radius $R_\star$ and luminosity of each star in various photometric bands (redshifted to $z_s = 0.725$) from the outputs of $\mathsf{FSPS}$. We do not simulate microlensing, and hence resolve only the stellar sizes of giant stars with $R_\star \sim 10^2$--$10^3\,R_\odot$ that approach the macrocaustic. We normalize the overall number of source stars to produce the measured surface brightness of the giant arc (which is preserved by lensing).

\begin{figure}[h]
  \begin{center}
  	\includegraphics[scale=0.62]{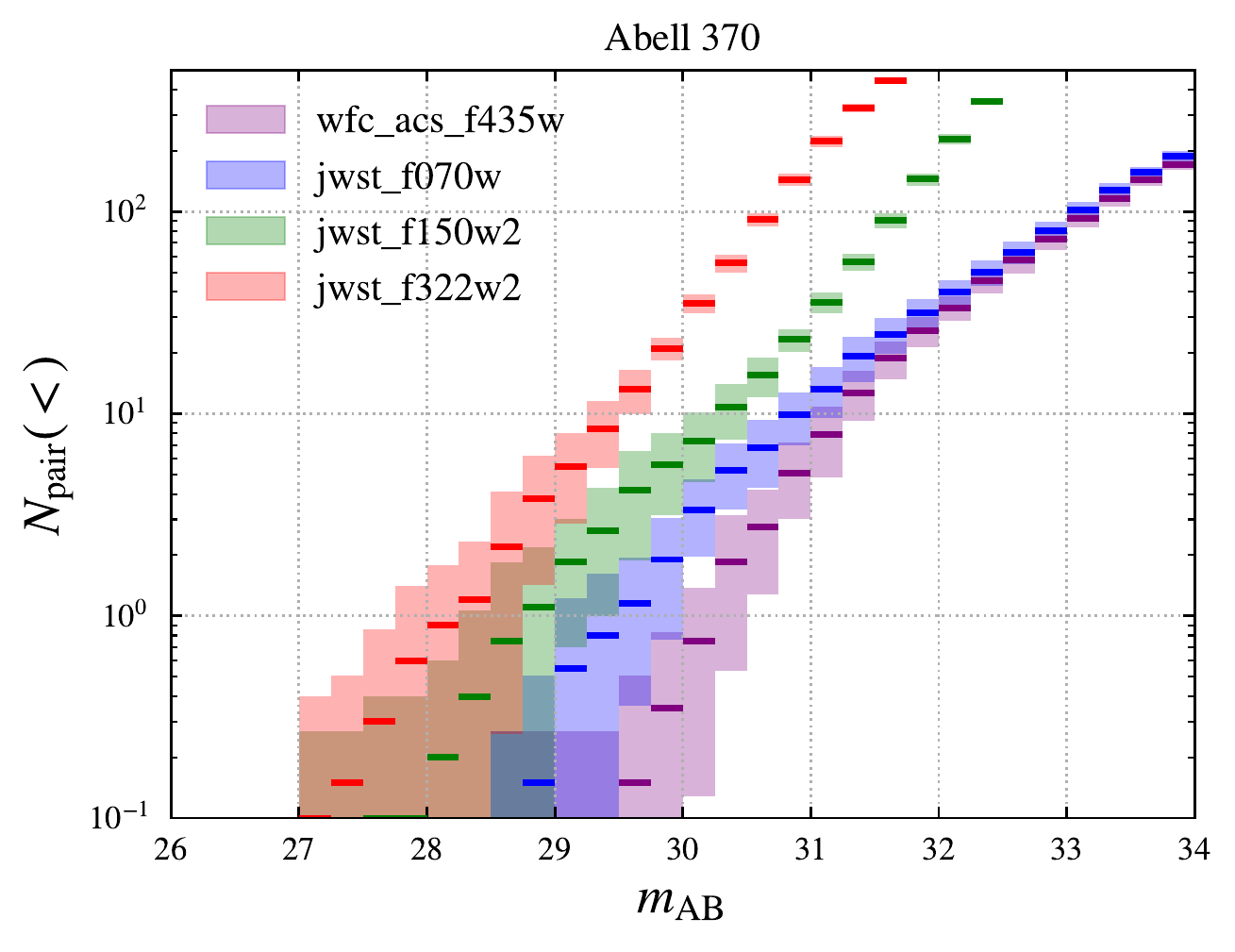}\\
    \includegraphics[scale=0.62]{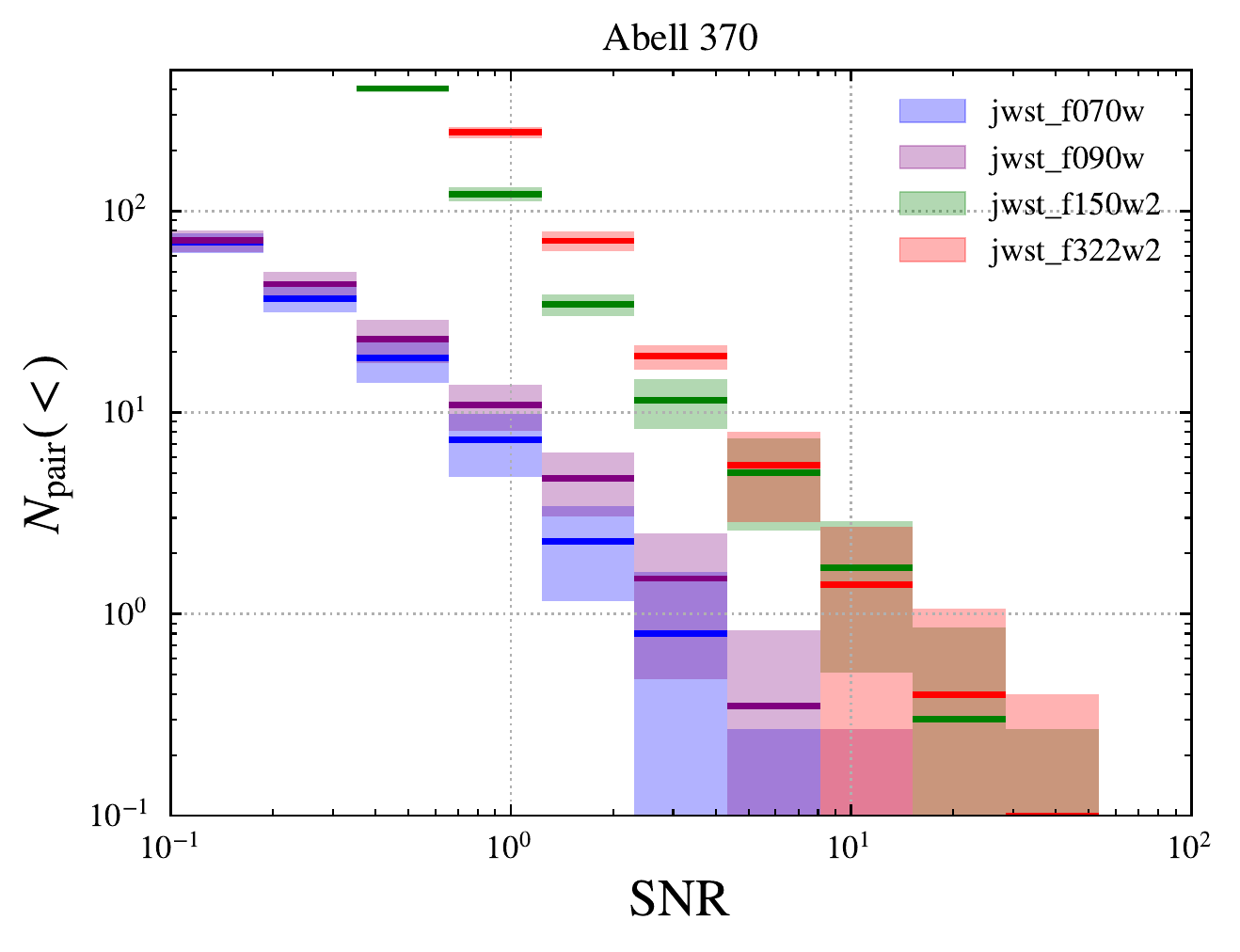}\\
    \includegraphics[scale=0.62]{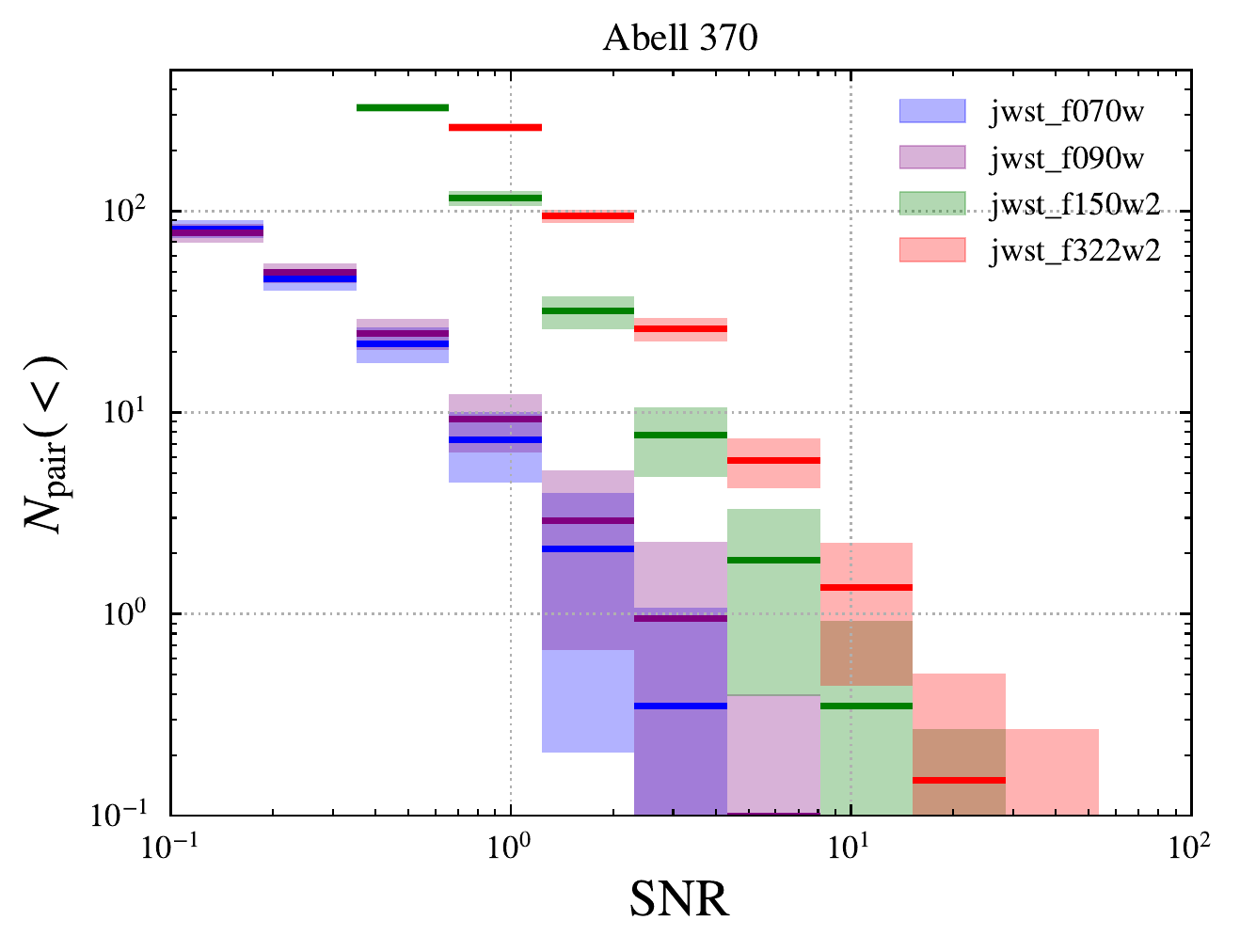}
    \caption{\label{fig:mag_dist} Expected number of highly magnified pairs of star images for Abell 370 assuming a simple fold model. {\it Upper panel}: Cumulative count of image pairs versus the maximum apparent magnitude. Results for a few selected {\it HST} and {\it JWST} wide filters are shown. {\it Middle panel}: Cumulative count of image pairs versus the minimum SNR of point-source detection with an exposure time of $6\, {\rm ks}$ with {\it JWST}, for four selected NIRCam wide filters. The mean metallicity is $\log(Z/Z_\odot)=-0.4$ according to our modeling. {\it Lower panel}: Same as middle panel but assuming $\log(Z/Z_\odot)=0.2$. In all panels, horizontal bars show the mean and shaded bands show the sample variance.}
  \end{center}
\end{figure}

We populate stars within an elongated $\sim 7\,{\rm kpc} \times 0.2\,{\rm kpc}$ region within the source galaxy (that the cluster caustic passes through); out of $\simeq 10^9$ stars, a few thousand have $L_{\rm bol} > 10^5\,L_\odot$. Individually detectable stars mainly belong to this ultra-luminous population. We compute the magnitudes of the (lensed) macro images for a smooth fold model, i.e., a model without dark matter substructure and without microlensing by intracluster stars. The upper panel of \reffig{mag_dist} shows the number of image pairs with apparent magnitudes brighter than a given value, and that lie within the $1'' \times 1''$ region at the intersection with the critical curve of the cluster, for the 4 filters indicated (one in {\it HST} and three in {\it JWST}). 

At fixed apparent AB magnitude, there are significantly more bright image pairs at $1$--$4\,\mu{\rm m}$ than at shorter wavelengths $\lesssim 1\,\mu{\rm m}$. Consistent with the observations in \refsec{stellar}, the brightest stars in the IR filters ({\sf f150w2} and {\sf f322w2}) and the bluer filter ({\sf f070w}) are red supergiants, and white/blue supergiants, respectively. Hot main-sequence stars dominate at shorter wavelengths ({\sf f435w}), but attain fainter apparent magnitudes than red supergiants achieve in the IR filters. We attribute this, at least in part, to stronger dust attenuation at shorter wavelengths.

The lower panel of \reffig{mag_dist} plots the cumulative number count of image pairs that can be detected in four filters of the {\it JWST} with a signal-to-noise ratio (SNR) higher than a given value, for $100$ min of integration time. We have calculated the SNRs specifically for the case of the giant arc in Abell 370, taking into account contaminating diffuse light including Zodical light, stray light from the Sun, the Galactic sky background, diffuse cluster light near the location of the caustic crossing stars, as well as light of the giant arc. For long exposures, point-source detection is limited by photon shot noise of the diffuse background. Based on archival {\it HST} images, we assume that the sum of the surface brightness from the giant arc and that from intracluster light at the caustic crossing location under examination is $S_{\rm arc+ICL}[{\rm mag}/{\rm arcsec}^2] = 22.5,\, 22.0,\, 21.5,\, 21.5$ at wavelengths $\lambda\,[\mu{\rm m}] = 0.7,\,0.9,\,1.5,\,3.2$, respectively (except that no {\it HST} observation is available at $\lambda = 3.2\,\mu$m). As for the sum of the Zodiacal light, stray light from the Sun, and the Galactic sky background, we assume a surface brightness $S_{\rm sbkg}[{\rm mag}/{\rm arcsec}^2] = 21.2,\, 21.1,\, 21.5,\, 22.4$ at wavelengths $\lambda\,[\mu{\rm m}] = 0.7,\,0.9,\,1.5,\,3.2$, respectively. In the two filters {\sf f150w2} and {\sf f322w2}, a few pairs of star images should already be detectable at $5\,\sigma$ with the assumed 100 minute exposure with {\it JWST}, and should rise to many tens of pairs with $\sim 20\,{\rm hr}$ of accumulated integration. The number of detectable stars, however, is significantly lower in shorter-wavelength filters at $\sim 0.7$--$1\,\mu{\rm m}$. We conclude that hot main-sequence stars are barely detectable without significant microlensing, even in {\it HST} bands. In contrast, many more red supergiants are detectable without any microlensing if observed at $1$--$4\,\mu{\rm m}$ with {\it JWST}. 

A spectroscopic study with {\it MUSE} found a high metallicity $12+\log({\rm O}/{\rm H})=8.88$ for the gaseous phase~\citep{patricio2018kinematics}, which can impact young stars that have formed recently. To check how uncertainty in the metallicity can impact the results, we perform another fit for the stellar population assuming a super-solar metallicity $\log(Z/Z_\odot)=0.2$. The results (lower panel of \reffig{mag_dist}) for the two infrared filters ${\sf f150w2}$ and ${\sf f322w2}$ suggest that the high-SNR tail (${\rm SNR} \gtrsim 10$) in the middle panel of \reffig{mag_dist} can decrease by a factor of few because ultra-luminous stars are reduced in a high-metallicity environment. Nevertheless, the number of fainter stars with ${\rm SNR} \simeq 1$--$3$ (which would be clearly detected with longer exposures) are not strongly affected.

Our results suggest that {\it JWST} can offer a better view of highly magnified stars than {\it HST} thanks to its sensitivity to red supergiants at infrared wavelengths. This conclusion for the arc in A370 is expected to remain valid for typical caustic-crossing systems.

Having prescriptions to generate random subhalos on the lens plane and random stars on the source plane, we are now able to simulate realizations of highly magnified stars in the vicinity of the critical curve, incorporating the effect of subhalo lensing. For each realization, we pixelize a $1.024'' \times 1.024''$ region centered at the smooth critical curve at a resolution of $8\,{\rm mas}\times 8\,{\rm mas}$. Detectable magnified stars have high luminosities, so it suffices to individually keep track of stars with $L_{\rm bol} > 10^4\,L_\odot$ and locate the pixels containing their macro images. See \refapp{raytrace} for details on how we numerically simulate lensing with substructure.

\subsection{Astrometric sensitivity to subhalo lensing}
\label{sec:results}

\begin{figure*}[t]
  \begin{center}
    \includegraphics[scale=0.41]{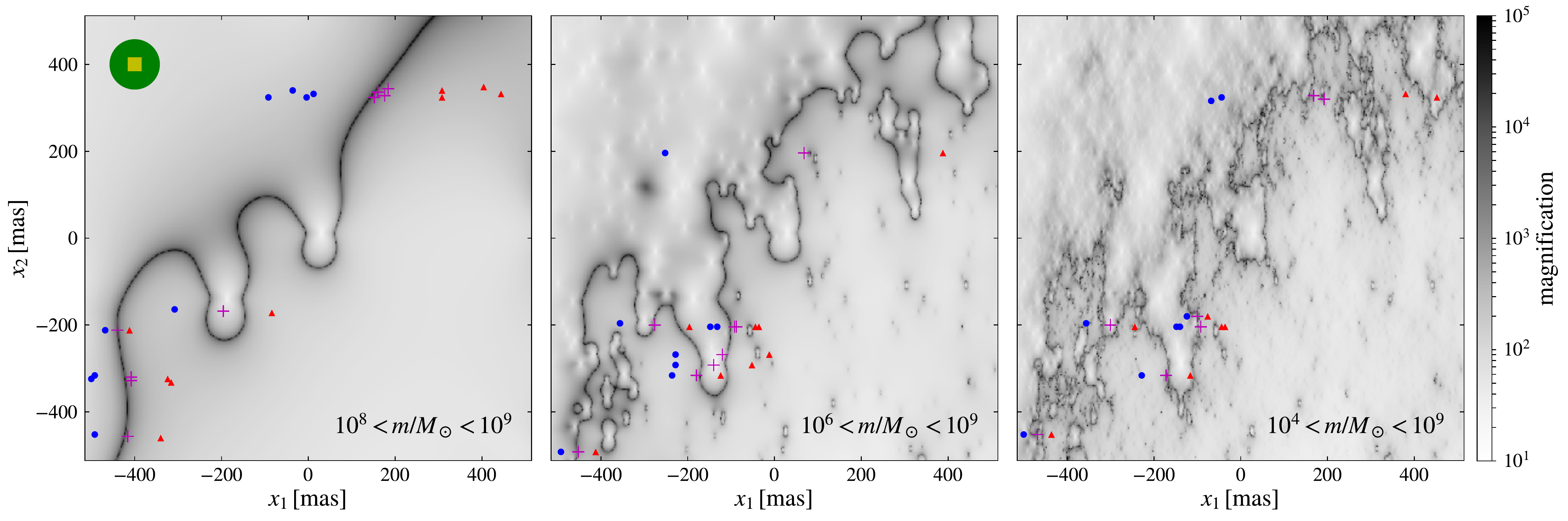}
    \caption{\label{fig:images} Simulated maps of the magnification factor (coded in gray scale) in a $1'' \times 1''$ region in the image plane, centered on the critical curve segment crossing the giant arc in Abell 370 shown in the zoomed-in panel of \reffig{a370}. The $x_1$ axis is along the direction of arc elongation (similar but not identical to the orientation of panels in \reffig{a370}). Image pair locations of magnified stars brighter than $m_{\rm AB} = 31$ in the {\it JWST} {\sf f150w2} filter are shown as blue dots and red triangles, with the midpoints marked as magenta crosses. From left to right, the same realization for source stars and subhalos is shown for three values of the minimum subhalo mass, as indicated. The yellow square in the first panel indicates the $\simeq 32\,{\rm mas}$ pixel size of {\it JWST}'s NIRCam in the short-wavelength camera, and the diffraction spread at $\lambda=1.5\,\mu$m is shown as the green disk. From left to right, the astrometric precision required to detect the effect of subhalos at a significance $\mathcal{S}=2$ after fitting the midpoints to a straight line (and a circle) is $\sigma_\theta = 59\,{\rm mas}$ ($64\,{\rm mas}$), $82\,{\rm mas}$ ($66\,{\rm mas}$), $85\,{\rm mas}$ ($64\,{\rm mas}$), respectively. In our simulation, the most luminous stars exhibit spatial clustering (toward the edges of the giant arc) because those short-lived stars should be associated with regions of active star formation within the host galaxy (notice the blue star-formation regions in the lower panel of \reffig{a370}.)}
  \end{center}
\end{figure*}

We now discuss lensing simulations computed for our fiducial parameters of the giant arc in Abell 370.

Based on a typical realization of subhalos and source stars in the giant arc, \reffig{images} exemplifies the subhalo lensing effect on the shape of the critical curves and the image positions of magnified stars. We choose the {\it JWST} ultra-wide filter {\sf f150w2} as an example. We select stars with both macro-images brighter than a threshold magnitude $m_{\rm AB} = 31$ (without microlensing), corresponding to a point-source detection SNR$\,\simeq 3$ ($5$) with $\sim 10\,{\rm hr}$ ($20\,{\rm hr}$) of integration.

\reffig{images} shows that subhalos strongly distort the smooth critical curve, create wiggles, and break it up into loops. To study the effects of subhalos of different masses, we set the maximum subhalo mass to $m = 10^9\,{\rm M}_\odot$ and successively include subhalos of smaller masses. Large subhalos $m \gtrsim 10^8\,{\rm M}_\odot$ (left panel) create features on large angular scales. Smaller subhalos ($10^6\,{\rm M}_\odot \lesssim m \lesssim 10^8\,{\rm M}_\odot$; middle panel), with a higher surface density, cause ubiquitous distortions on angular scales $\lesssim 100\,{\rm mas}$. Very small subhalos $10^4\,{\rm M}_\odot \lesssim m \lesssim 10^6\,{\rm M}_\odot$ imprint distortions on very fine angular scales $\lesssim 10\,{\rm mas}$, which are difficult to resolve with current or forthcoming instruments.

 As seen in \reffig{images}, most detectable image pairs of magnified stars are located within $\sim 0.1''$ of the cluster critical curve. Because of the perturbations by subhalos, the astrometric midpoints of the image pairs do not align along a smooth curve. To test the null hypothesis that the simple fold model can account for the image positions, we fit the midpoints to a straight line using orthogonal regression~\citep{deming1943statistical}. Under the simplifying assumption that all image pairs have an equal, uncorrelated astrometric precision, we quantify the statistical significance of the departure from the null hypothesis with the quantity,
\ba
\label{eq:calS}
\mathcal{S} = \left[ \frac{2}{\sigma^2_\theta} \sum^{N_p}_{i=1}\,s^2_i - (N_p - p) \right]^{1/2}.
\ea
Here, $N_p$ is the total number of image pairs with astrometric measurements, $i = 1,2,\cdots, N_p$ enumerates the image pairs, and $s_i$ is the perpendicular residual from the best-fit critical curve to each midpoint. For the most simple approximation to the critical curve, a straight line, the number of free parameters is $p=2$.
We use $\sigma_\theta$ to denote the standard deviation of astrometric errors of image positions along each Cartesian axis. The astrometric standard deviation of the midpoint of two images is then $\sigma_\theta/\sqrt{2}$. For the realization shown in \reffig{images}, an astrometric precision of $\sigma_\theta \simeq 60$--$80\,{\rm mas}$ is required to reject the null hypothesis at $2\sigma$ ($\mathcal{S} = 2$). 

Despite their low number density, subhalos of larger masses $m \gtrsim 10^{10}\,M_\odot$ (which may be hosting dwarf or large galaxies) can accidentally lie close to the critical curves. Alternatively, a group of less massive subhalos clustered together by chance can collectively generate a significant lensing perturbation. Acting at intermediate distances (say $\gtrsim 1''$), these effects can induce a local curvature in the smooth critical curve. If the optical counterpart of a large subhalo is measured, its contribution to lensing can be modeled and corrected, but this will generally still leave uncorrected perturbations from relatively massive subhalos. In order to relax the null hypothesis to account for this possibility, we modify our model fitting the midpoints to a partial circle, for which $p=3$ in \refeq{calS}. However, a very small best-fit radius of curvature suggests the presence of dark subhalos (if not accompanied by a dwarf galaxy in the lensing cluster) in the mass range we are interested to detect. For this reason, we impose a minimum radius of curvature $300\,{\rm mas}$ to prevent falsely excluding evidence for substructure.

Our analysis assumes that image pairs can be correctly identified and that astrometric accuracy is not worsened by blending and crowding. In practice, the identification of image pairs may not be easy, but is aided by the known and fixed direction of elongation along which all pairs should be split.

\begin{figure*}[t]
  \begin{center}
    \includegraphics[scale=0.6]{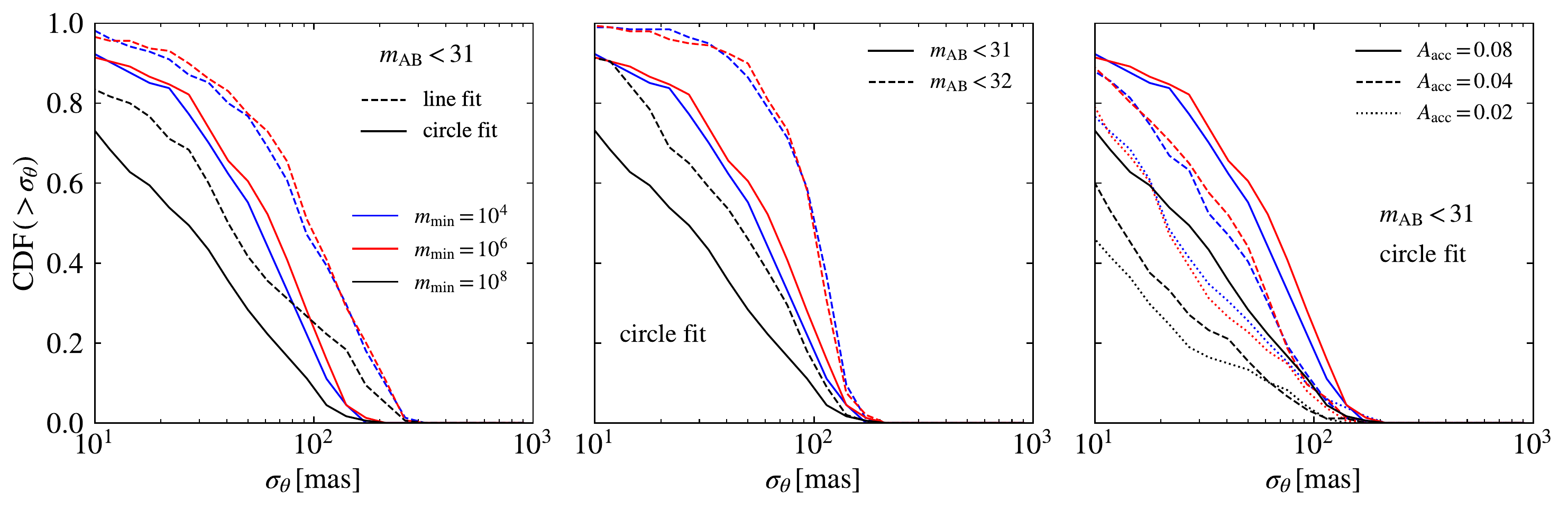}
    \caption{\label{fig:sigd} Cumulative distribution functions for the required astrometric precision $\sigma_\theta$ to rule out the fold model at $2\sigma$ ($\mathcal{S}=2$) for the giant arc in Abell 370. We select image pairs brighter than a threshold magnitude in the {\it JWST}'s {\sf f150w2} filter. Different curve colors correspond to minimum subhalo masses $\log(m_{\rm min}/M_\odot) = 4,\,6,\,8$. Each curve is computed based on 200 random realizations. {\it Left}: Comparison between fitting the midpoints of the image pairs to a straight line and fitting them to a circle. {\it Middle}: Comparison between different magnitude limits $m_{\rm AB} < 31,\,32$ for image detection. {\it Right:} Effect of varying the normalization of the subhalo abundance $A_{\rm acc}$ (\refeq{dndlnmacc}) from the fiducial value $A_{\rm acc} = 0.08$. (Due to pixelization at $8" \times 8"$ resolution in our simulation, $\sigma_\theta$ has an artificial minimum at the level of $8\,$mas; this is however unimportant compared to a physical signature $\sigma_\theta \sim $ tens of mas.)}
  \end{center}
\end{figure*}

To statistically quantify the required astrometric precision for rejecting the null hypothesis, we simulate a large number of random realizations of subhalos and source stars, and perform both the line fit and the circle fit for each realization as in the case of \reffig{images}. \reffig{sigd} shows that the cumulative distribution of the required $\sigma_\theta$ has a median at $\sim 30$--$80$ mas. Larger values of $\sigma_\theta$ are required when subhalos smaller than $10^8\,M_\odot$ are added, but this saturates for $m_{\rm min} \lesssim 10^6\,M_\odot$. Theoretically speaking, too many low-mass subhalos can limit the maximum magnification factor and therefore suppress the overall number of detectable bright images. Our simulation results show this effect when the minimum subhalo mass is decreased from $10^8\,M_\odot$ to $10^6\,M_\odot$, but we find no evidence that this trend continues to $m \lesssim 10^6\,M_\odot$. We therefore conclude that the astrometric test with magnified stars is most sensitive to subhalos in the mass range $10^6\,M_\odot \lesssim m \lesssim 10^8\,M_\odot$. More than six orders of magnitude below the mass scale of the host halo, these are beyond the reach of the state-of-the-art N-body simulations of halo formation. Since star formation should be quenched due to the hot intracluster environment~\citep{2014MNRAS.440.1934T}, we expect these subhalos to be non-luminous systems. Constraints on their abundance should be of great interest as a probe to the dark matter.

The left panel of \reffig{sigd} shows that allowing the smooth critical curve to have a small curvature (radius of curvature $> 0.3''$) can reduce but not fully eliminate the astrometric residuals. As demonstrated by \reffig{images}, perturbations by subhalos shift the midpoints of image pairs incoherently, in a way that cannot be described by a slight curvature of the critical curve.

The middle panel of \reffig{sigd} shows that detecting more caustic crossing stars enhances the astrometric sensitivity to subhalo lensing, because an increased number of tracer stars allows for a denser sampling of the critical curve across the width of the giant arc. 

In the right panel of \reffig{sigd}, we vary the normalization $A_{\rm acc}$ of the subhalo mass function to assess the sensitivity of the subhalo perturbations to their abundance. We find that if the subhalo surface number density is reduced by a factor of four compared to our fiducial model, there is still a large probability that the lensing imprints of subhalos can be revealed along the giant arc with an astrometric precision of $\gtrsim 30\,{\rm mas}$.

Despite the large uncertainty in the subhalo abundance after an extrapolation down to $m \simeq 10^4$--$10^8\,M_\odot$, the results suggest that the phenomenon we study here should be observationally detectable in standard CDM. These observations can constrain several properties of the subhalos like concentration and tidal radius, which will need more detailed studies.

\section{Discussion and conclusions}
\label{sec:concl}

We have proposed a novel method to probe the subhalo contents of cluster dark matter halos in the low-mass regime $10^6$--$10^8\,M_\odot$. Although substructure on these mass scales have been numerically studied for galaxy-sized halos~\citep{2008Natur.454..735D}, state-of-the-art N-body simulations of individual galaxy clusters have not been able to resolve down to that level, corresponding to an impressive subhalo-to-host mass ratio $\sim 10^{-7}$--$10^{-9}$. Probing this wide dynamic range tests structure formation on sub-galactic scales and explores modifications of dark matter from the CDM paradigm.

The proposed method uses astrometry of highly magnified image pairs of stars in caustic-straddling giant arcs to seek subhalo-induced departures from the symmetric positions expected in a smooth fold. The method is different from previously proposed methods to detect sub-structure induced astrometric anomalies in lensed quasars~\citep[see e.g.][]{Metcalf:2002ra}. In particular, we take the advantage that these astrometric perturbations are strongly amplified when subhalos lie close in projection to a critical curve. The identification of image pairs of source stars is aided by their property of aligning along a unique direction of elongation. Microlensing by intracluster stars is not expected to appreciably shift image positions but causes flux variability. This information can help identify image pairs, but one must properly account for the additional flux fluctuation due to microlensing of unresolved stars within a single pixel~ \citep[see e.g.][]{Tuntsov:2004gj}.

Cluster member galaxies in subhalos with mass $m \gtrsim 10^{10}\,M_\odot$ can also cause curvature on the critical curve. However, distortions from these subhalos should be coherent over patches around the critical curve as large as $\sim 1''$, and are distinguishable from the signatures of smaller subhalos. The lensing influence of nearby galaxies can be modeled and corrected if their optical counterparts are detected.

As a specific example, we have studied the giant arc of Abell 370 in the vicinity of one caustic-crossing location, by simulating random realizations of subhalos based on a realistic model of halo substructure. We found that with  images of $\simeq 5$ to $10$ stars with astrometric precisions $\sigma_\theta \simeq 20$--$100\,{\rm mas}$, the astrometric perturbations by subhalos of masses $m \simeq 10^6$--$10^8\,M_\odot$ can most often be detected.

The forth-coming {\it JWST} offers improved prospects to detect enough lensed stars with the required astrometry to probe subhalos compared to {\it HST}, owing to increased number of bright stars, particularly supergiants, that can be detected at $1-4\,\mu$m in typical star-forming galaxies at $z \simeq 1$. For the giant arc of Abell 370, we forecast that $\sim 5$ to $10$ stars down to $m_{\rm AB} < 31$ can be measured with $\sim 10\,{\rm hr}$ integrations with the widest {\it JWST} filters. Current and forthcoming ground-based giant telescopes can reach higher angular resolution with adaptive optics in the near-infrared and may also be promising for this purpose.

Finding new caustic-straddling galaxies that can probe dark matter substructure would be highly beneficial. The best targets are low-redshift, star-forming galaxies (to maximize the number of detectable stars) with a local convergence $\kappa_0$ close to unity and a small shear gradient $|\bfd|$ (to maximize magnification of the stellar flux and the substructure perturbations), and a large arc width that allows sampling a larger portion of the critical curve.

Our study has considered only subhalos orbiting within the lensing cluster, ignoring intervening field halos that are randomly projected along the line of sight. The projected abundance of intervening halos, which have better chances of surviving tidal disruption than subhalos inside clusters, can dominate over that of cluster subhalos by a factor of few in the mass range of our interest~\citep{Despali:2017ksx}, and have been considered important in applications of strong lensing reconstruction~\citep{Birrer:2016xku, 2017ApJ...836..141M} and quasar flux ratio tests~\citep{metcalf2005importance, metcalf2005testing, xu2012effects, xu2015well}. Simulating lensing by intervening halos requires ray tracing with multiple lens planes, which is beyond the scope of this work. Intervening halos should be included in future work and are expected to enhance and make more easily detectable the effect of substructure we have discussed.

Intracluster globular clusters (GCs)~\citep{white1987globular, west1995intracluster} observed in nearby clusters~\citep{2010Sci...328..334L, peng2011hst, west2011globular, Alamo-Martinez:2013aaa, d2016extended, lee2016globular} may also generate astrometric perturbations through lensing. Having typical masses $10^4$--$10^6\,M_\odot$, these are on average less massive than halos at the lower end ($\sim 10^6\,M_\odot$) of the interesting subhalo mass range, but should be more efficient lenses due to their high concentrations. However, their surface density of $\lesssim {\rm few}\times 10^{-2}\,{\rm kpc}^{-2}$ at the typical impact parameter of the critical curve~\citep{ramos2017intra} is two orders of magnitude smaller than for subhalos of $m \gtrsim 10^6\,M_\odot$ (see \reffig{shabun}). Therefore, intracluster GCs are unlikely to be a major source of confusion for detecting dark matter subhalos. 

Deep images at high resolution are likely to reveal star clusters and compact star-forming associates in the source galaxy. Near a caustic, their physical size $l$ limits the maximum magnification factor $\mu \lesssim [2\,(l/D_S)\,|d\,\sin\alpha|]^{-1/2}/[2(1-\kappa_0)]$~\citep{Venumadhav:2017ttg}, or $\mu \lesssim 600\,(l/{\rm pc})^{-1/2}$ in the case of Abell 370, at which the elongated image has a length $\sim 50\,{\rm mas}\,(l/{\rm pc})^{1/2}$. Depending on the compactness, these structures may be under-resolved or marginally resolved. They may be more easily detectable as they are typically brighter than individual stars, and they are not susceptible to microlensing. These sources may be used as substitutes for individual stars in the astrometric test.

Finally, surface brightness irregularities exist in the source galaxy, which include HII regions, open clusters, features of spiral arms or dust lanes. Those can cause complications in the detection of individual stars. In multi-epoch observations, those should be distinguishable from individual stars based on flux variabilities, since extended source sizes should quench microlensing. On the other hand, surface brightness irregularities themselves can be used to probe subhalos, in which case a variety of techniques developed for other applications (e.g. strong lensing of sub-mm galaxies) are applicable. Further studies are warranted to assess the impact of surface brightness irregularities near caustics.

Given the above considerations, it would be valuable to have mock telescope images created for ultra-bright stars on top of a realistic source galaxy surface brightness profile and with the effect of confusion accounted for. More sophisticated treatment at the level of realistic data analysis is warranted for future work.

To conclude, observing the effect of mass clumps near cluster critical curves is a powerful probe to constrain the physical nature of the dark matter. It is a promising method to robustly test theories of warm dark matter or ultralight boson dark matter that make different predictions for the substructure inside dark matter halos.

\begin{acknowledgments}
\acknowledgments

The authors thank Hsiao-Wen Chen, Brenda Frye, Simon White, Rogier Windhorst, Barak Zackay, Matias Zaldarriaga for inspiring discussions. JM thanks the Institute for Advanced Study for their support during visits.

We made use of the lens model for Abell 370 implemented in {\sf Lenstool}. We modeled the giant arc based on archival {\it HST} images. We computed point-source detection SNRs for caustic crossing stars using the {\it JWST} Exposure Time Calculator.

LD is supported at the Institute for Advanced Study by NASA through
Einstein Postdoctoral Fellowship grant number PF5-160135 awarded by
the Chandra X-ray Center, which is operated by the Smithsonian
Astrophysical Observatory for NASA under contract NAS8-03060. TV and AAK acknowledge support from the Schmidt Fellowship. TV is also supported by the W.M. Keck Foundation Fund. JM was partially supported by Spanish MINECO grants AYA2015-71091-P and MDM-2014-0369.

\end{acknowledgments}

\appendix

\section{Subhalo abundance}
\label{app:shabund}

In this Appendix, we detail how we model the subhalo content of the host galaxy cluster expected in the standard $\Lambda$CDM cosmology.

N-body simulations suggest that dark matter halos are well described by the Navarro-Frenk-White (NFW) density profile~\citep{1996ApJ...462..563N}. In galaxy clusters, the NFW profile appears to fit the sum of the dark matter and the baryonic matter~\citep{newman2013density}. As a function of the halocentric distance $R$, the density is given by $\rho(R) = \rho_{\rm crit}\,\delta_c/(R/R_s)/(1+R/R_s)^2$. Here $R_s$ is the scale radius, $\rho_{\rm crit}(z) = 3 H^2(z)/(8 \pi G)$ is the critical density, and the characteristic overdensity $\delta_c$ depends on the concentration parameter $C_{200} = R_{200}/R_s$ through $\delta_c =(200/3)\,(C^3_{200}/f(C_{200}))$, where $f(x)\equiv \ln(1+x)-x/(1+x)$, and $R_{200}$ is the radius within which the mean density is 200 times the cosmic mean. A commonly adopted characteristic mass is the total mass enclosed within $R_{200}$. For field halos, we use the mean concentration-mass-redshift relation $C_{200} = \bar{C}_{200}(M_{200}, z)$ found by \cite{ludlow2016mass}.   

In hierarchical structure formation, a cluster-sized halo acquires substructure by accreting smaller halos~\citep{kravtsov2004tumultuous}. Each subhalo has an initial mass $m_{\rm acc}$ at the time of accretion and subsequently loses mass to tidal stripping. While the {\it evolved} subhalo mass function depends substantially on the host mass, the {\it unevolved} mass function in terms of $m_{\rm acc}$ appears universal~\citep{2008MNRAS.386.2135G}. This is the picture of ``unbiased accretion'', in which the unevolved subhalo specific mass function spatially traces the host density profile~\citep{Han:2015pua} and is parametrized as \refeq{dndlnmacc}, which becomes inaccurate only in the inner part $R/R_{200} \lesssim 0.1$~\citep{2011MNRAS.412...49W, Jiang:2014zfa}.

The tidal stripping ratio $m/m_{\rm acc}$ is subject to large scatter because it is sensitive to the complete subhalo orbital history. In particular, subhalos on eccentric orbits suffer from strong tidal stripping and tidal shocking during pericenter passages~\citep{Hayashi:2002qv}. On average, a strong scaling of the tidal stripping ratio with the halocentric distance $m/m_{\rm acc} \propto R^{\beta}$ with $\beta \sim 1$ is seen. Following \cite{Han:2015pua}, we assign a probability for bound mass $m$ given an infall mass $m_{\rm acc}$,
\ba
\label{eq:mfrommacc}
&& dP\left(m|m_{\rm acc}, R\right) = (1- f_s)\,\delta_D(m)\,dm + f_s\,\mathcal{N}\left( \ln\frac{m}{m_{\rm acc}}, \ln\mu_{\rm ts}(R), \sigma_{\rm ts} \right)\,d\ln m.
\ea
The first term accounts for a finite fraction $(1-f_s)$ of complete tidal disruption, while the second term describes a random draw for $m/m_{\rm acc}$ from a log-normal distribution with a mean $\mu_{\rm ts}(R) = \mu_\star(R/R_{200})^\beta$ and a constant variance $\sigma_{\rm ts}$. To satisfy physical constraints, the log-normal distribution is truncated to $m/m_{\rm acc} < 1$. For a cluster-sized host halo, we use fiducial values $f_s = 0.56$, $\mu_\star = 0.34$, $\beta = 1.0$ and $\sigma_{\rm ts} = 1.1$ as suggested by simulations~\citep{Han:2015pua}.

\section{Subhalo lens model}
\label{app:shlen}

Many studies found that substructure halos have systematically higher concentrations than comparable field halos and are subject to larger scatter~\citep{2001MNRAS.321..559B, 2007ApJ...667..859D, 2015PhRvD..92l3508B}. This may be explained by tidal stripping, or may be related to biased halo formation within the proto-cluster. \cite{2017MNRAS.466.4974M} quantified the enhancement of subhalo concentration in simulated Milky Way-sized halos, which increases as $\propto R^{-1/2}$ toward smaller halocentric distances. Guided by these findings, we match \cite{2017MNRAS.466.4974M} with the prescription for subhalo concentration given in \refeq{shc200}. We introduce this prescription as a simple, reasonable model, with the caveat that it is unclear how the results of \cite{2017MNRAS.466.4974M} may be extrapolated to cluster-sized host halos and our knowledge about subhalo internal structure is in general poor. 

With our fiducial model set up, subhalos are generated in the following way: for a given infall mass $m_{\rm acc}$ at a given halocentric distance $R$, we draw the bound mass $m$ according to \refeq{mfrommacc}, compute $c_{200}$ from \refeq{shc200}, compute $r_t$ from \refeq{instidal}, and finally tune $m_{200}$ and $r_t$ such that the total mass equals to $m$. 

We now present the analytical results for the lensing effect of a subhalo with our chosen density profile, the truncated NFW profile in \refeq{shrho}. The gravitational potential gradient is
\ba
\frac{d\Phi_{\rm sh}(r)}{dr} = \frac{G m_{200}}{f(c_{200})\,r^2}\,H_1\left( \frac{r}{r_s}, \tau \right),
\ea
where $\tau \equiv r_t/r_s$, and we have defined the auxiliary function
\ba
H_1(t, \tau) & \equiv & \frac{\tau^2}{2\,(1+t)\,(1+\tau^2)^2}\,\left[ -2\,t\,(1+\tau^2) + 4\,\tau\,(1+t)\,\tan^{-1}\frac{t}{\tau} \right. \en
&& \left. + 2\,(1+t)\,(\tau^2 - 1)\,\ln[(1+t)\,\tau] - (1+t)\,(\tau^2 - 1)\,\ln(t^2 + \tau^2) \right] > 0 ~. 
\ea
The gravitational potential is given by
\ba
\Phi_{\rm sh}(r) = - \frac{G\,m_{200}}{f(c_{200})\,r_s}\,H_2\left( \frac{r}{r_s}, \tau \right),
\ea
where $H_2(t, \tau)$ is a second auxiliary function,
\ba
H_2(t, \tau) & \equiv & \frac{\tau}{2\,t\,(1+\tau^2)^2}\,\left[ -2\,\left( 2\,\tau^2 + t\,(\tau^2 - 1) \right)\,\tan^{-1}\frac{\tau}{t} + \tau\,\left( 2\,\pi\,\tau - 2\,t\,(1+\tau^2)\,\ln\left(1+\frac1t\right) \right.\right.\en
&& \left.\left. - 2\,(1+t)\,(\tau^2 - 1)\,\ln\frac{t}{1+t} + 2\,(\tau^2 - 1)\,\ln\tau + (1 + 2\,t - \tau^2)\,\ln\left( 1 + \frac{\tau^2}{t^2} \right) \right) \right] > 0 ~.
\ea
The lensing potential is given by
\ba
\psi_{\rm sh}(b) = - \frac{D_{LS}}{D_L\,D_S}\, \frac{4\,G\,m_{200}}{c^2\,f(c_{200})}\,\int^{+\infty}_0\,d\eta\,H_2\left( \sqrt{\eta^2 + \left( \frac{b}{r_s} \right)^2}, \tau \right) ~,
\ea
where $c$ is the speed of light in vacuum, and $D_L$, $D_S$ and $D_{LS}$ are the angular diameter distances to the lens, to the source, and from the lens to the source, respectively. The impact parameter $b$ is the related to the angular impact parameter through $b = D_L\,|\bfx|$, where $\bfx$ is the angular displacement vector on the lens plane.

The lensing deflection at an angle $\bfx$ away from the center of the subhalo is given by
\ba
\bfalp_{\rm sh}(\bfx) = \frac{\bfx}{|\bfx|}\, \frac{D_{LS}}{D_S}\,\frac{4\,G\,m_{200}}{c^2\,f(c_{200})\,r_s}\,S\left( \frac{D_L\,|\bfx|}{r_s},\, \tau \right),
\ea
where we introduce a dimensionless function
\ba
S(\xi,\,\tau) \equiv \int^{+\infty}_0\,d\eta\,H_1(\sqrt{\eta^2 + \xi^2}, \tau)\,\frac{\xi}{(\eta^2 + \xi^2)^{3/2}} ~.
\ea
Similarly, the lensing convergence is given by
\ba
\label{eq:subh_kappa}
\kappa_{\rm sh}(\bfx) = \frac12\,\frac{D_{LS}\,D_L}{D_S}\,\frac{4\,G\,m_{200}}{c^2\,f(c_{200})\,r^2_s}\,K\left( \frac{D_L\,|\bfx|}{r_s} , \tau \right).
\ea
There, a second dimensionless function reads
\ba
K(\xi,\,\tau) & \equiv & \frac{\partial\,S(\xi,\,\tau)}{\partial \xi} + \frac{S(\xi,\,\tau)}{\xi} = \int^{+\infty}_0\,\frac{\xi^2\,d\eta}{(\eta^2 + \xi^2)^{2}}\,\left[ H_{1,1}(\sqrt{\eta^2 + \xi^2}, \tau) + \left( \frac{2\,\eta^2}{\xi^2} - 1 \right)\, \frac{H_1(\sqrt{\eta^2 + \xi^2}, \tau)}{\sqrt{\eta^2 + \xi^2}} \right],
\ea
with our notation for partial derivatives $H_{1,1}(t,\tau) \equiv \partial_t\,H_1(t,\tau)$.

Finally, the two components of the lensing shear can be computed from $\gamma_{\rm sh, 1} = (1/2)\,(\partial_{x_1} \alpha_{\rm sh, 1} - \partial_{x_2} \alpha_{\rm sh, 2})$ and $\gamma_{\rm sh, 2} = \partial_{x_1} \alpha_{\rm sh, 2}$, and are given by
\ba
\label{eq:subh_gamma}
\left[\begin{array}{c}
\gamma_{\rm sh, 1}(\bfx) \\
\gamma_{\rm sh, 2}(\bfx) \\
\end{array}\right]
& = & - \frac12\,\frac{D_{LS}\,D_L}{D_S}\,\frac{4\,G\,m_{200}}{c^2\,f(c_{200})\,r^2_s}\,G\left( \frac{D_L\,|\bfx|}{r_s} , \tau \right)\, \left[\begin{array}{c}
\cos\,2\phi \\
\sin\,2\phi \\
\end{array}\right].
\ea
Here $\phi$ is the orientation angle of the vector $\bfx$ on the lens plane, and we have introduced another dimensionless function,
\ba
G(\xi,\,\tau) & \equiv & \frac{S(\xi,\,\tau)}{\xi} - \frac{\partial\,S(\xi,\,\tau)}{\partial \xi} = \int^{+\infty}_0\,\frac{\xi^2\,d\eta}{(\eta^2 + \xi^2)^{2}}\,\left[ \frac{3\,H_1(\sqrt{\eta^2 + \xi^2}, \tau)}{\sqrt{\eta^2 + \xi^2}}- H_{1,1}(\sqrt{\eta^2 + \xi^2}, \tau) \right].
\ea

\section{Simulating lensing}
\label{app:raytrace}

In this Appendix, we briefly discuss how we simulate lensing of stars in the vicinity of a cluster lensing caustic. Each $1.024'' \times 1.024''$ field of view centered on the critical curve is divided into $N_{\rm side} \times N_{\rm side}$ square pixels with $N_{\rm side}=128$. Each pixel has a size $8\,{\rm mas} \times 8\,{\rm mas}$. 

Each pixel is sampled with the technique of inverse ray shooting~\citep{kayser1986astrophysical, schneider1987gravitational, 1990PhDT.......180W, wambsganss1999gravitational} on an adaptively refined grid, which is designed to ensure good accuracy near critical curves. Starting with the entire pixel, we shoot rays at the four vertices and at the center, and compute deflection, the lensing Jacobian matrix, and the magnification factor. If the magnification varies substantially (by more than 20\%) across the pixel, or if it changes sign (suggesting crossing of a critical curve), the pixel is further divided into four equal sub-pixels, and rays are shot at their vertices and centers. The algorithm proceeds recursively until either the magnification becomes sufficiently uniform within a sub-pixel, or a maximum number of recursions $N_{\rm recur}=8$ have been performed.

The source plane is sampled using a technique similar to the polygon mapping method~\citep{mediavilla2006fast, mediavilla2011new}. Rays shot on the adaptively refined grid on the image plane are mapped onto the source plane, where they define the vertices of polygons that cover the source plane. Special sub-pixels that straddle critical curves on the image plane are further divided into parts on either side of the critical curve and are then mapped to polygons on the source plane. The magnification factor is then interpolated from the values at the vertices. For a source of finite size, such as a stellar disk, we calculate the fraction contained within each polygon (either the source is contained within a single polygon or it straddles several neighboring polygons) and compute the flux using the interpolated magnification. For each polygon on the source plane, we efficiently search through all source stars by organizing them into a hierarchical tree according to their positions~\citep{wambsganss1999gravitational}. This method enables to correctly compute the magnifications of small but finite sources by adaptively shooting a realistic number of rays.
\bigskip

\bibliographystyle{aasjournal}
\bibliography{reference_subhalo,reference_subhalo_more}

\begin{thebibliography}{}
\expandafter\ifx\csname natexlab\endcsname\relax\def\natexlab#1{#1}\fi

\bibitem[{Abell(1958)}]{abell1958distribution}
Abell, G.~O. 1958, The Astrophysical Journal Supplement Series, 3, 211

\bibitem[{Ade {et~al.}(2016)Ade, Aghanim, Arnaud, Ashdown, Aumont, Baccigalupi,
  Banday, Barreiro, Bartlett, Bartolo, {et~al.}}]{ade2016planck}
Ade, P.~A., Aghanim, N., Arnaud, M., {et~al.} 2016, Astronomy \& Astrophysics,
  594, A13

\bibitem[{Alamo-Martínez {et~al.}(2013)Alamo-Martínez, Blakeslee, Jee,
  Côté, Ferrarese, González-Lópezlira, Jordán, Meurer, Peng, \&
  West}]{Alamo-Martinez:2013aaa}
Alamo-Martínez, K.~A., Blakeslee, J.~P., Jee, M.~J., {et~al.} 2013, Astrophys.
  J., 775, 20

\bibitem[{Asadi {et~al.}(2017)Asadi, Zackrisson, \&
  Freeland}]{asadi2017probing}
Asadi, S., Zackrisson, E., \& Freeland, E. 2017, Monthly Notices of the Royal
  Astronomical Society, 472, 129

\bibitem[{Baltz {et~al.}(2009)Baltz, Marshall, \& Oguri}]{Baltz:2007vq}
Baltz, E.~A., Marshall, P., \& Oguri, M. 2009, JCAP, 0901, 015

\bibitem[{{Bartels} \& {Ando}(2015)}]{2015PhRvD..92l3508B}
{Bartels}, R., \& {Ando}, S. 2015, \prd, 92, 123508

\bibitem[{B{\'e}zecourt {et~al.}(1999)B{\'e}zecourt, Ellis, Soucail, \&
  Kneib}]{bezecourt1999lensed}
B{\'e}zecourt, J., Ellis, R.~S., Soucail, G., \& Kneib, J. 1999, Astron.
  Astrophys., 351, 433

\bibitem[{B{\'e}zecourt {et~al.}(1998)B{\'e}zecourt, Kneib, Soucail, \&
  Ebbels}]{bezecourt1998lensed}
B{\'e}zecourt, J., Kneib, J., Soucail, G., \& Ebbels, T. 1998, arXiv preprint
  astro-ph/9810199

\bibitem[{Binney \& Tremaine(1987)}]{binney1987galactic}
Binney, J., \& Tremaine, S. 1987, Galactic Dynamics, Princeton Univ,  Press
  Princeton, NJ;

\bibitem[{{Birrer} {et~al.}(2017){Birrer}, {Amara}, \&
  {Refregier}}]{2017JCAP...05..037B}
{Birrer}, S., {Amara}, A., \& {Refregier}, A. 2017, JCAP, 5, 037

\bibitem[{Birrer {et~al.}(2017)Birrer, Welschen, Amara, \&
  Refregier}]{Birrer:2016xku}
Birrer, S., Welschen, C., Amara, A., \& Refregier, A. 2017, JCAP, 1704, 049

\bibitem[{Blumenthal {et~al.}(1984)Blumenthal, Faber, Primack, \&
  Rees}]{blumenthal1984formation}
Blumenthal, G.~R., Faber, S., Primack, J.~R., \& Rees, M.~J. 1984

\bibitem[{Blumenthal {et~al.}(1982)Blumenthal, Pagels, \&
  Primack}]{blumenthal1982galaxy}
Blumenthal, G.~R., Pagels, H., \& Primack, J.~R. 1982, Nature, 299, 37

\bibitem[{Bode {et~al.}(2001)Bode, Ostriker, \& Turok}]{bode2001halo}
Bode, P., Ostriker, J.~P., \& Turok, N. 2001, The Astrophysical Journal, 556,
  93

\bibitem[{{Bullock}(2010)}]{2010arXiv1009.4505B}
{Bullock}, J.~S. 2010, ArXiv e-prints, arXiv:1009.4505

\bibitem[{{Bullock} {et~al.}(2001){Bullock}, {Kolatt}, {Sigad}, {Somerville},
  {Kravtsov}, {Klypin}, {Primack}, \& {Dekel}}]{2001MNRAS.321..559B}
{Bullock}, J.~S., {Kolatt}, T.~S., {Sigad}, Y., {et~al.} 2001, \mnras, 321, 559

\bibitem[{Carlberg(2009)}]{carlberg2009star}
Carlberg, R. 2009, The Astrophysical Journal Letters, 705, L223

\bibitem[{Carr(1975)}]{Carr:1975qj}
Carr, B.~J. 1975, Astrophys. J., 201, 1

\bibitem[{Carr \& Hawking(1974)}]{carr1974black}
Carr, B.~J., \& Hawking, S.~W. 1974, Monthly Notices of the Royal Astronomical
  Society, 168, 399

\bibitem[{Charlot \& Fall(2000)}]{charlot2000simple}
Charlot, S., \& Fall, S.~M. 2000, The Astrophysical Journal, 539, 718

\bibitem[{Colin {et~al.}(2000)Colin, Avila-Reese, \&
  Valenzuela}]{colin2000substructure}
Colin, P., Avila-Reese, V., \& Valenzuela, O. 2000, The Astrophysical Journal,
  542, 622

\bibitem[{Conroy \& Gunn(2010)}]{conroy2010propagation}
Conroy, C., \& Gunn, J.~E. 2010, The Astrophysical Journal, 712, 833

\bibitem[{Conroy {et~al.}(2009)Conroy, Gunn, \& White}]{conroy2009propagation}
Conroy, C., Gunn, J.~E., \& White, M. 2009, The Astrophysical Journal, 699, 486

\bibitem[{Contini {et~al.}(2014)Contini, De~Lucia, Villalobos, \&
  Borgani}]{Contini:2013wha}
Contini, E., De~Lucia, G., Villalobos, ., \& Borgani, S. 2014, Mon. Not. Roy.
  Astron. Soc., 437, 3787

\bibitem[{Cooper {et~al.}(2015)Cooper, Gao, Guo, Frenk, Jenkins, Springel, \&
  White}]{Cooper:2014nwa}
Cooper, A.~P., Gao, L., Guo, Q., {et~al.} 2015, Mon. Not. Roy. Astron. Soc.,
  451, 2703

\bibitem[{Covone {et~al.}(2006)Covone, Kneib, Soucail, Richard, Jullo, \&
  Ebeling}]{covone2006vimos}
Covone, G., Kneib, J.-P., Soucail, G., {et~al.} 2006, Astronomy \&
  Astrophysics, 456, 409

\bibitem[{Cyr-Racine {et~al.}(2016)Cyr-Racine, Moustakas, Keeton, Sigurdson, \&
  Gilman}]{Cyr-Racine:2015jwa}
Cyr-Racine, F.-Y., Moustakas, L.~A., Keeton, C.~R., Sigurdson, K., \& Gilman,
  D.~A. 2016, Phys. Rev., D94, 043505

\bibitem[{Dalal \& Kochanek(2002)}]{dalal2002direct}
Dalal, N., \& Kochanek, C. 2002, The Astrophysical Journal, 572, 25

\bibitem[{{Davis} {et~al.}(1985){Davis}, {Efstathiou}, {Frenk}, \&
  {White}}]{1985ApJ...292..371D}
{Davis}, M., {Efstathiou}, G., {Frenk}, C.~S., \& {White}, S.~D.~M. 1985, \apj,
  292, 371

\bibitem[{{Davis} {et~al.}(1981){Davis}, {Lecar}, {Pryor}, \&
  {Witten}}]{1981ApJ...250..423D}
{Davis}, M., {Lecar}, M., {Pryor}, C., \& {Witten}, E. 1981, \apj, 250, 423

\bibitem[{Deming(1943)}]{deming1943statistical}
Deming, W.~E. 1943

\bibitem[{Despali {et~al.}(2017)Despali, Vegetti, White, Giocoli, \& van~den
  Bosch}]{Despali:2017ksx}
Despali, G., Vegetti, S., White, S. D.~M., Giocoli, C., \& van~den Bosch, F.~C.
  2017, arXiv:1710.05029

\bibitem[{Diego {et~al.}(2017)}]{Diego:2017drh}
Diego, J.~M., {et~al.} 2017, arXiv:1706.10281

\bibitem[{{Diemand} {et~al.}(2007){Diemand}, {Kuhlen}, \&
  {Madau}}]{2007ApJ...667..859D}
{Diemand}, J., {Kuhlen}, M., \& {Madau}, P. 2007, \apj, 667, 859

\bibitem[{{Diemand} {et~al.}(2008){Diemand}, {Kuhlen}, {Madau}, {Zemp},
  {Moore}, {Potter}, \& {Stadel}}]{2008Natur.454..735D}
{Diemand}, J., {Kuhlen}, M., {Madau}, P., {et~al.} 2008, \nat, 454, 735

\bibitem[{Diemand {et~al.}(2005)Diemand, Moore, \& Stadel}]{diemand2005earth}
Diemand, J., Moore, B., \& Stadel, J. 2005, Nature, 433, 389

\bibitem[{D’Abrusco {et~al.}(2016)D’Abrusco, Cantiello, Paolillo, Pota,
  Napolitano, Limatola, Spavone, Grado, Iodice, Capaccioli,
  {et~al.}}]{d2016extended}
D’Abrusco, R., Cantiello, M., Paolillo, M., {et~al.} 2016, The Astrophysical
  Journal Letters, 819, L31

\bibitem[{Fort {et~al.}(1988)Fort, Prieur, Mathez, Mellier, \&
  Soucail}]{fort1988faint}
Fort, B., Prieur, J., Mathez, G., Mellier, Y., \& Soucail, G. 1988, Astronomy
  and Astrophysics, 200, L17

\bibitem[{{Giocoli} {et~al.}(2008){Giocoli}, {Tormen}, \& {van den
  Bosch}}]{2008MNRAS.386.2135G}
{Giocoli}, C., {Tormen}, G., \& {van den Bosch}, F.~C. 2008, \mnras, 386, 2135

\bibitem[{Goodman(2000)}]{goodman2000repulsive}
Goodman, J. 2000, New Astronomy, 5, 103

\bibitem[{Han {et~al.}(2016)Han, Cole, Frenk, \& Jing}]{Han:2015pua}
Han, J., Cole, S., Frenk, C.~S., \& Jing, Y. 2016, Mon. Not. Roy. Astron. Soc.,
  457, 1208

\bibitem[{Hayashi {et~al.}(2003)Hayashi, Navarro, Taylor, Stadel, \&
  Quinn}]{Hayashi:2002qv}
Hayashi, E., Navarro, J.~F., Taylor, J.~E., Stadel, J., \& Quinn, T.~R. 2003,
  Astrophys. J., 584, 541

\bibitem[{Hezaveh {et~al.}(2013)Hezaveh, Dalal, Holder, Kuhlen, Marrone,
  Murray, \& Vieira}]{hezaveh2013dark}
Hezaveh, Y., Dalal, N., Holder, G., {et~al.} 2013, The Astrophysical Journal,
  767, 9

\bibitem[{Hezaveh {et~al.}(2016)Hezaveh, Dalal, Marrone, Mao, Morningstar, Wen,
  Blandford, Carlstrom, Fassnacht, Holder, {et~al.}}]{hezaveh2016detection}
Hezaveh, Y.~D., Dalal, N., Marrone, D.~P., {et~al.} 2016, The Astrophysical
  Journal, 823, 37

\bibitem[{{Hoekstra} {et~al.}(2013){Hoekstra}, {Bartelmann}, {Dahle}, {Israel},
  {Limousin}, \& {Meneghetti}}]{2013SSRv..177...75H}
{Hoekstra}, H., {Bartelmann}, M., {Dahle}, H., {et~al.} 2013, \ssr, 177, 75

\bibitem[{Hu {et~al.}(2000)Hu, Barkana, \& Gruzinov}]{PhysRevLett.85.1158}
Hu, W., Barkana, R., \& Gruzinov, A. 2000, Phys. Rev. Lett., 85, 1158

\bibitem[{Hui {et~al.}(2017)Hui, Ostriker, Tremaine, \&
  Witten}]{PhysRevD.95.043541}
Hui, L., Ostriker, J.~P., Tremaine, S., \& Witten, E. 2017, Phys. Rev. D, 95,
  043541

\bibitem[{Ibata {et~al.}(2002)Ibata, Lewis, Irwin, \&
  Quinn}]{ibata2002uncovering}
Ibata, R., Lewis, G., Irwin, M., \& Quinn, T. 2002, Monthly Notices of the
  Royal Astronomical Society, 332, 915

\bibitem[{{Iocco} {et~al.}(2015){Iocco}, {Pato}, \&
  {Bertone}}]{2015NatPh..11..245I}
{Iocco}, F., {Pato}, M., \& {Bertone}, G. 2015, Nature Physics, 11, 245

\bibitem[{Jiang {et~al.}(2015)Jiang, Cole, Sawala, \& Frenk}]{Jiang:2014zfa}
Jiang, L., Cole, S., Sawala, T., \& Frenk, C.~S. 2015, Mon. Not. Roy. Astron.
  Soc., 448, 1674

\bibitem[{Johnston {et~al.}(2002)Johnston, Spergel, \&
  Haydn}]{johnston2002lumpy}
Johnston, K.~V., Spergel, D.~N., \& Haydn, C. 2002, The Astrophysical Journal,
  570, 656

\bibitem[{{Jullo} \& {Kneib}(2009)}]{2009MNRAS.395.1319J}
{Jullo}, E., \& {Kneib}, J.-P. 2009, \mnras, 395, 1319

\bibitem[{{Katz} {et~al.}(1986){Katz}, {Balbus}, \&
  {Paczynski}}]{1986ApJ...306....2K}
{Katz}, N., {Balbus}, S., \& {Paczynski}, B. 1986, \apj, 306, 2

\bibitem[{Kayser {et~al.}(1986)Kayser, Refsdal, \&
  Stabell}]{kayser1986astrophysical}
Kayser, R., Refsdal, S., \& Stabell, R. 1986, Astronomy and Astrophysics, 166,
  36

\bibitem[{{Kelly} {et~al.}(2017){Kelly}, {Diego}, {Rodney}, {Kaiser},
  {Broadhurst}, {Zitrin}, {Treu}, {Perez-Gonzalez}, {Morishita}, {Jauzac},
  {Selsing}, {Oguri}, {Pueyo}, {Ross}, {Filippenko}, {Smith}, {Hjorth},
  {Cenko}, {Wang}, {Howell}, {Richard}, {Frye}, {Jha}, {Foley}, {Norman},
  {Bradac}, {Zheng}, {Brammer}, {Molino Benito}, {Cava}, {Christensen}, {de
  Mink}, {Graur}, {Grillo}, {Kawamata}, {Kneib}, {Matheson}, {McCully},
  {Nonino}, {Perez-Fournon}, {Riess}, {Rosati}, {Borello Schmidt}, {Sharon}, \&
  {Weiner}}]{2017arXiv170610279K}
{Kelly}, P.~L., {Diego}, J.~M., {Rodney}, S., {et~al.} 2017, ArXiv e-prints,
  arXiv:1706.10279

\bibitem[{Kneib {et~al.}(1993)Kneib, Mellier, Fort, \&
  Mathez}]{kneib1993distribution}
Kneib, J., Mellier, Y., Fort, B., \& Mathez, G. 1993, Astronomy and
  Astrophysics, 273, 367

\bibitem[{Kravtsov {et~al.}(2004)Kravtsov, Gnedin, \&
  Klypin}]{kravtsov2004tumultuous}
Kravtsov, A.~V., Gnedin, O.~Y., \& Klypin, A.~A. 2004, The Astrophysical
  Journal, 609, 482

\bibitem[{{Lagattuta} {et~al.}(2017){Lagattuta}, {Richard}, {Cl{\'e}ment},
  {Mahler}, {Patr{\'{\i}}cio}, {Pell{\'o}}, {Soucail}, {Schmidt}, {Wisotzki},
  {Martinez}, \& {Bina}}]{2017MNRAS.469.3946L}
{Lagattuta}, D.~J., {Richard}, J., {Cl{\'e}ment}, B., {et~al.} 2017, \mnras,
  469, 3946

\bibitem[{Lee \& Jang(2016)}]{lee2016globular}
Lee, M.~G., \& Jang, I.~S. 2016, The Astrophysical Journal, 831, 108

\bibitem[{{Lee} {et~al.}(2010){Lee}, {Park}, \& {Hwang}}]{2010Sci...328..334L}
{Lee}, M.~G., {Park}, H.~S., \& {Hwang}, H.~S. 2010, Science, 328, 334

\bibitem[{{Lin} \& {Mohr}(2004)}]{2004ApJ...617..879L}
{Lin}, Y.-T., \& {Mohr}, J.~J. 2004, \apj, 617, 879

\bibitem[{Ludlow {et~al.}(2016)Ludlow, Bose, Angulo, Wang, Hellwing, Navarro,
  Cole, \& Frenk}]{ludlow2016mass}
Ludlow, A.~D., Bose, S., Angulo, R.~E., {et~al.} 2016, Monthly Notices of the
  Royal Astronomical Society, 460, 1214

\bibitem[{Lynds \& Petrosian(1986)}]{lynds1986giant}
Lynds, R., \& Petrosian, V. 1986, in Bulletin of the American Astronomical
  Society, Vol.~18, 1014

\bibitem[{Lynds \& Petrosian(1989)}]{lynds1989luminous}
Lynds, R., \& Petrosian, V. 1989, The Astrophysical Journal, 336, 1

\bibitem[{{Mandelbaum}(2015)}]{2015IAUS..311...86M}
{Mandelbaum}, R. 2015, in IAU Symposium, Vol. 311, Galaxy Masses as Constraints
  of Formation Models, ed. M.~{Cappellari} \& S.~{Courteau}, 86--95

\bibitem[{Mao \& Schneider(1998)}]{Mao:1997ek}
Mao, S.-d., \& Schneider, P. 1998, Mon. Not. Roy. Astron. Soc., 295, 587

\bibitem[{Martel {et~al.}(2012)Martel, Barai, \& Brito}]{Martel:2012ue}
Martel, H., Barai, P., \& Brito, W. 2012, Astrophys. J., 757, 48

\bibitem[{{McCully} {et~al.}(2017){McCully}, {Keeton}, {Wong}, \&
  {Zabludoff}}]{2017ApJ...836..141M}
{McCully}, C., {Keeton}, C.~R., {Wong}, K.~C., \& {Zabludoff}, A.~I. 2017,
  \apj, 836, 141

\bibitem[{Mediavilla {et~al.}(2011)Mediavilla, Mediavilla, Mu{\~n}oz, Ariza,
  Lopez, Gonzalez-Morcillo, \& Jimenez-Vicente}]{mediavilla2011new}
Mediavilla, E., Mediavilla, T., Mu{\~n}oz, J., {et~al.} 2011, The Astrophysical
  Journal, 741, 42

\bibitem[{Mediavilla {et~al.}(2006)Mediavilla, Mu{\~n}oz, Lopez, Mediavilla,
  Abajas, Gonzalez-Morcillo, \& Gil-Merino}]{mediavilla2006fast}
Mediavilla, E., Mu{\~n}oz, J., Lopez, P., {et~al.} 2006, The Astrophysical
  Journal, 653, 942

\bibitem[{Mellier {et~al.}(1991)Mellier, Fort, Soucail, Mathez, \&
  Cailloux}]{mellier1991spectroscopy}
Mellier, Y., Fort, B., Soucail, G., Mathez, G., \& Cailloux, M. 1991, The
  Astrophysical Journal, 380, 334

\bibitem[{Mellier {et~al.}(1988)Mellier, Soucail, Fort, \&
  Mathez}]{mellier1988photometry}
Mellier, Y., Soucail, G., Fort, B., \& Mathez, G. 1988, Astronomy and
  Astrophysics, 199, 13

\bibitem[{Meszaros(1974)}]{meszaros1974behaviour}
Meszaros, P. 1974, Astronomy and Astrophysics, 37, 225

\bibitem[{Metcalf(2002)}]{Metcalf:2002ra}
Metcalf, R.~B. 2002, Astrophys. J., 580, 696

\bibitem[{Metcalf(2005{\natexlab{a}})}]{metcalf2005importance}
---. 2005{\natexlab{a}}, The Astrophysical Journal, 629, 673

\bibitem[{Metcalf(2005{\natexlab{b}})}]{metcalf2005testing}
---. 2005{\natexlab{b}}, The Astrophysical Journal, 622, 72

\bibitem[{Metcalf \& Madau(2001)}]{metcalf2001compound}
Metcalf, R.~B., \& Madau, P. 2001, The Astrophysical Journal, 563, 9

\bibitem[{Miralda-Escude(1991)}]{miralda1991magnification}
Miralda-Escude, J. 1991, The Astrophysical Journal, 379, 94

\bibitem[{Mo {et~al.}(2010)Mo, Van~den Bosch, \& White}]{mo2010galaxy}
Mo, H., Van~den Bosch, F., \& White, S. 2010, Galaxy formation and evolution
  (Cambridge University Press)

\bibitem[{{Molin{\'e}} {et~al.}(2017){Molin{\'e}}, {S{\'a}nchez-Conde},
  {Palomares-Ruiz}, \& {Prada}}]{2017MNRAS.466.4974M}
{Molin{\'e}}, {\'A}., {S{\'a}nchez-Conde}, M.~A., {Palomares-Ruiz}, S., \&
  {Prada}, F. 2017, \mnras, 466, 4974

\bibitem[{{Navarro} {et~al.}(1996){Navarro}, {Frenk}, \&
  {White}}]{1996ApJ...462..563N}
{Navarro}, J.~F., {Frenk}, C.~S., \& {White}, S.~D.~M. 1996, \apj, 462, 563

\bibitem[{Newman {et~al.}(2013)Newman, Treu, Ellis, Sand, Nipoti, Richard, \&
  Jullo}]{newman2013density}
Newman, A.~B., Treu, T., Ellis, R.~S., {et~al.} 2013, The Astrophysical
  Journal, 765, 24

\bibitem[{{Nierenberg} {et~al.}(2017){Nierenberg}, {Treu}, {Brammer}, {Peter},
  {Fassnacht}, {Keeton}, {Kochanek}, {Schmidt}, {Sluse}, \&
  {Wright}}]{2017MNRAS.471.2224N}
{Nierenberg}, A.~M., {Treu}, T., {Brammer}, G., {et~al.} 2017, \mnras, 471,
  2224

\bibitem[{Oguri {et~al.}(2017)Oguri, Diego, Kaiser, Kelly, \&
  Broadhurst}]{Oguri:2017ock}
Oguri, M., Diego, J.~M., Kaiser, N., Kelly, P.~L., \& Broadhurst, T. 2017,
  arXiv:1710.00148

\bibitem[{{Okamoto} {et~al.}(2008){Okamoto}, {Gao}, \&
  {Theuns}}]{2008MNRAS.390..920O}
{Okamoto}, T., {Gao}, L., \& {Theuns}, T. 2008, \mnras, 390, 920

\bibitem[{Patricio {et~al.}(2018)Patricio, Richard, Carton, Contini, Epinat,
  Brinchmann, Schmidt, Krajnovic, Bouche, Weilbacher,
  {et~al.}}]{patricio2018kinematics}
Patricio, V., Richard, J., Carton, D., {et~al.} 2018, arXiv preprint
  arXiv:1802.08451

\bibitem[{Peebles(2000)}]{peebles2000fluid}
Peebles, P. 2000, The Astrophysical Journal Letters, 534, L127

\bibitem[{Peng {et~al.}(2011)Peng, Ferguson, Goudfrooij, Hammer, Lucey, Marzke,
  Puzia, Carter, Balcells, Bridges, {et~al.}}]{peng2011hst}
Peng, E.~W., Ferguson, H.~C., Goudfrooij, P., {et~al.} 2011, The Astrophysical
  Journal, 730, 23

\bibitem[{Press \& Schechter(1974)}]{press1974formation}
Press, W.~H., \& Schechter, P. 1974, The Astrophysical Journal, 187, 425

\bibitem[{Profumo {et~al.}(2006)Profumo, Sigurdson, \&
  Kamionkowski}]{PhysRevLett.97.031301}
Profumo, S., Sigurdson, K., \& Kamionkowski, M. 2006, Phys. Rev. Lett., 97,
  031301

\bibitem[{Ramos-Almendares {et~al.}(2017)Ramos-Almendares, Abadi, Muriel, \&
  Coenda}]{ramos2017intra}
Ramos-Almendares, F., Abadi, M.~G., Muriel, H., \& Coenda, V. 2017, arXiv
  preprint arXiv:1712.05410

\bibitem[{{Richard} {et~al.}(2010){Richard}, {Kneib}, {Limousin}, {Edge}, \&
  {Jullo}}]{2010MNRAS.402L..44R}
{Richard}, J., {Kneib}, J.-P., {Limousin}, M., {Edge}, A., \& {Jullo}, E. 2010,
  \mnras, 402, L44

\bibitem[{Schneider {et~al.}(1992)Schneider, Ehlers, \&
  Falco}]{schneider1992gravitational}
Schneider, P., Ehlers, J., \& Falco, E. 1992, Gravitational Lenses
  Gravitational Lenses, XIV, 560 pp. 112 figs,  Springer-Verlag Berlin
  Heidelberg New York. Also Astronomy and Astrophysics Library

\bibitem[{Schneider {et~al.}(1999)Schneider, Ehlers, \&
  Falco}]{schneider1999gravitational}
---. 1999, Gravitational Lenses, Astronomy and Astrophysics Library (Springer)

\bibitem[{Schneider \& Weiss(1987)}]{schneider1987gravitational}
Schneider, P., \& Weiss, A. 1987, Astronomy and Astrophysics, 171, 49

\bibitem[{Sin(1994)}]{PhysRevD.50.3650}
Sin, S.-J. 1994, Phys. Rev. D, 50, 3650

\bibitem[{Smail {et~al.}(1995)Smail, Dressler, Kneib, Ellis, Couch, Sharples,
  \& Oemler~Jr}]{smail1995hst}
Smail, I., Dressler, A., Kneib, J.-P., {et~al.} 1995, arXiv preprint
  astro-ph/9503063

\bibitem[{{Sofue} \& {Rubin}(2001)}]{2001ARA&A..39..137S}
{Sofue}, Y., \& {Rubin}, V. 2001, \araa, 39, 137

\bibitem[{Soucail {et~al.}(1987)Soucail, Fort, Mellier, \&
  Picat}]{soucail1987blue}
Soucail, G., Fort, B., Mellier, Y., \& Picat, J. 1987, Astronomy and
  Astrophysics, 172, L14

\bibitem[{Soucail {et~al.}(1988)Soucail, Mellier, Fort, Mathez, \&
  Cailloux}]{soucail1988giant}
Soucail, G., Mellier, Y., Fort, B., Mathez, G., \& Cailloux, M. 1988, Astronomy
  and Astrophysics, 191, L19

\bibitem[{{Taranu} {et~al.}(2014){Taranu}, {Hudson}, {Balogh}, {Smith},
  {Power}, {Oman}, \& {Krane}}]{2014MNRAS.440.1934T}
{Taranu}, D.~S., {Hudson}, M.~J., {Balogh}, M.~L., {et~al.} 2014, \mnras, 440,
  1934

\bibitem[{{Treu}(2010)}]{2010ARA&A..48...87T}
{Treu}, T. 2010, \araa, 48, 87

\bibitem[{Tuntsov {et~al.}(2004)Tuntsov, Lewis, Ibata, \&
  Kneib}]{Tuntsov:2004gj}
Tuntsov, A.~V., Lewis, G.~F., Ibata, R.~A., \& Kneib, J.-P. 2004, Mon. Not.
  Roy. Astron. Soc., 353, 853

\bibitem[{Turner(1983)}]{PhysRevD.28.1243}
Turner, M.~S. 1983, Phys. Rev. D, 28, 1243

\bibitem[{Umetsu {et~al.}(2011)Umetsu, Broadhurst, Zitrin, Medezinski, \&
  Hsu}]{umetsu2011cluster}
Umetsu, K., Broadhurst, T., Zitrin, A., Medezinski, E., \& Hsu, L.-Y. 2011, The
  Astrophysical Journal, 729, 127

\bibitem[{{van den Bosch} {et~al.}(2018){van den Bosch}, {Ogiya}, {Hahn}, \&
  {Burkert}}]{2018MNRAS.474.3043V}
{van den Bosch}, F.~C., {Ogiya}, G., {Hahn}, O., \& {Burkert}, A. 2018, \mnras,
  474, 3043

\bibitem[{Vegetti {et~al.}(2010)Vegetti, Koopmans, Bolton, Treu, \&
  Gavazzi}]{vegetti2010detection}
Vegetti, S., Koopmans, L., Bolton, A., Treu, T., \& Gavazzi, R. 2010, Monthly
  Notices of the Royal Astronomical Society, 408, 1969

\bibitem[{Vegetti {et~al.}(2012)Vegetti, Lagattuta, McKean, Auger, Fassnacht,
  \& Koopmans}]{vegetti2012gravitational}
Vegetti, S., Lagattuta, D., McKean, J., {et~al.} 2012, Nature, 481, 341

\bibitem[{Venumadhav {et~al.}(2017)Venumadhav, Dai, \&
  Miralda-Escud¨¦}]{Venumadhav:2017ttg}
Venumadhav, T., Dai, L., \& Miralda-Escud¨¦, J. 2017, Astrophys. J., 850, 49

\bibitem[{Viel {et~al.}(2005)Viel, Lesgourgues, Haehnelt, Matarrese, \&
  Riotto}]{PhysRevD.71.063534}
Viel, M., Lesgourgues, J., Haehnelt, M.~G., Matarrese, S., \& Riotto, A. 2005,
  Phys. Rev. D, 71, 063534

\bibitem[{{Wambsganss}(1990)}]{1990PhDT.......180W}
{Wambsganss}, J. 1990, PhD thesis, Thesis Ludwig-Maximilians-Univ., Munich
  (Germany, F.~R.).~Fakult{\"a}t f{\"u}r Physik., (1990)

\bibitem[{Wambsganss(1999)}]{wambsganss1999gravitational}
Wambsganss, J. 1999, Journal of Computational and Applied Mathematics, 109, 353

\bibitem[{West {et~al.}(1995)West, C{\^o}t{\'e}, Jones, Forman, \&
  Marzke}]{west1995intracluster}
West, M.~J., C{\^o}t{\'e}, P., Jones, C., Forman, W., \& Marzke, R.~O. 1995,
  The Astrophysical Journal Letters, 453, L77

\bibitem[{West {et~al.}(2011)West, Jord{\'a}n, Blakeslee, C{\^o}t{\'e}, Gregg,
  Takamiya, \& Marzke}]{west2011globular}
West, M.~J., Jord{\'a}n, A., Blakeslee, J.~P., {et~al.} 2011, Astronomy \&
  Astrophysics, 528, A115

\bibitem[{{Wetzel}(2011)}]{2011MNRAS.412...49W}
{Wetzel}, A.~R. 2011, \mnras, 412, 49

\bibitem[{White(1987)}]{white1987globular}
White, R.~E. 1987, Monthly Notices of the Royal Astronomical Society, 227, 185

\bibitem[{White \& Rees(1978)}]{white1978core}
White, S.~D., \& Rees, M.~J. 1978, Monthly Notices of the Royal Astronomical
  Society, 183, 341

\bibitem[{Windhorst {et~al.}(2018)}]{Windhorst:2018wft}
Windhorst, R.~A., {et~al.} 2018, Astrophys. J. Suppl., 234, 41

\bibitem[{Xu {et~al.}(2012)Xu, Mao, Cooper, Gao, Frenk, Angulo, \&
  Helly}]{xu2012effects}
Xu, D., Mao, S., Cooper, A.~P., {et~al.} 2012, Monthly Notices of the Royal
  Astronomical Society, 421, 2553

\bibitem[{Xu {et~al.}(2015)Xu, Sluse, Gao, Wang, Frenk, Mao, Schneider, \&
  Springel}]{xu2015well}
Xu, D., Sluse, D., Gao, L., {et~al.} 2015, Monthly Notices of the Royal
  Astronomical Society, 447, 3189

\bibitem[{{Zibetti} {et~al.}(2005){Zibetti}, {White}, {Schneider}, \&
  {Brinkmann}}]{2005MNRAS.358..949Z}
{Zibetti}, S., {White}, S.~D.~M., {Schneider}, D.~P., \& {Brinkmann}, J. 2005,
  \mnras, 358, 949

\bibitem[{{Zwicky}(1951)}]{1951PASP...63...61Z}
{Zwicky}, F. 1951, \pasp, 63, 61

\end{thebibliography}

\end{document}